\documentclass[a4paper,pra,onecolumn,superscriptaddress]{revtex4}
\usepackage[english]{babel}
\usepackage[utf8]{inputenc}
\usepackage{graphicx,epsfig,color}
\usepackage{latexsym}
\usepackage{amsfonts} 
\usepackage{amsmath}
\usepackage{textcomp}
\usepackage{hyperref}
\usepackage{placeins}
\usepackage{booktabs}

\newcommand{\Tr}{\mathrm{Tr}}
\def\ket|#1>{| #1 \rangle}
\def\bra<#1|{\langle #1 |}
\def\<{\langle}
\def\>{\rangle}
\def\{{\lbrace}
\def\}{\rbrace}
\def\({\left(}
\def\){\right)}
\def\[{\left[}
\def\]{\right]}
\def\beq{\begin{equation}}
\def\eeq{\end{equation}}

\def\opt{\text{opt}}
\def\mut{\text{mut}}
\def\cont{\text{cont}}
\def\tot{\text{tot}}

\def\bbar#1{\bar{\bar #1}}

\begin{document}

\title{Link representation of the entanglement entropies for all bipartitions }

\author{Sudipto Singha Roy}
\affiliation{Instituto de Física Teórica UAM/CSIC, Universidad
  Autónoma de Madrid, Cantoblanco, Madrid, Spain}

\author{Silvia N. Santalla}
\affiliation{Dep.~de Física and Grupo Interdisciplinar de Sistemas
  Complejos (GISC), Universidad Carlos III de Madrid, Spain}

\author{Germán Sierra}
\affiliation{Instituto de Física Teórica UAM/CSIC, Universidad
  Autónoma de Madrid, Cantoblanco, Madrid, Spain}

\author{Javier Rodríguez-Laguna}
\affiliation{Dep.~de Física Fundamental, UNED, Madrid, Spain}

\begin{abstract}
  We have recently shown that the entanglement entropy of any bipartition of a quantum state can be approximated as the sum of certain link strengths connecting internal and external sites. The representation is useful to unveil the geometry associated with the entanglement structure of a quantum many-body state which may occasionally differ from the one suggested by the Hamiltonian of the system. Yet, the obtention of these entanglement links is a complex mathematical problem. In this work, we address this issue and propose several approximation techniques for matrix product states, free fermionic states, or in cases in which contiguous blocks are specially relevant. Along with this, we discuss the accuracy of the approximation for different types of states and partitions.  Finally, we employ the link representation to discuss two different physical systems: the spin-1/2 long-range XXZ chain and the spin-1 bilinear biquadratic chain.
\end{abstract}

\date{\today}

\maketitle


\section{Introduction}

Entanglement is a key feature for the new developments in quantum
physics. It constitutes the main resource for quantum technologies
\cite{Horodecki_09}, characterizes the different phases of quantum
matter \cite{Osterloh_02,Osborne_02}, and has been proposed as the
building block of the fabric of space-time through the holographic
principle and tensor networks
\cite{Maldacena_99,Ryu_06,Vidal_07,Raamsdonk_10,Swingle_12,Cao_17,Hyatt_17}.
This last conjecture derives from the so-called {\em area law}, which
states that the entanglement entropy of a block of a low-lying energy
eigenstate of a local Hamiltonian is proportional to the measure to
its boundary \cite{Amico_08,Sredniki_93,Eisert_10,Wolf_08}, sometimes
presenting logarithmic corrections
\cite{Holzhey_94,Vidal_03,Calabrese_04,Calabrese_09}. Interestingly,
there are relevant exceptions to the area law, such as the {\em
  rainbow state}
\cite{Vitagliano_10,Ramirez_14,Ramirez_15,Laguna_16,Laguna_17,Tonni_18,
  Alba_19,Samos_19,MacCormack_19,Samos_20}. In this case, despite the
locality of the Hamiltonian, the ground state establishes long-range
bonds between opposite sites of a chain, suggesting the possibility
that the geometric structure described by the entanglement differs
from the one associated to the Hamiltonian. Thus, given a quantum
state, we are naturally led to the question: what is the geometry
associated to its entanglement structure? A tentative answer to that
question was provided by our group in \cite{SinghaRoy_20}, and we will
build upon that work in order to provide a deeper insight.

The main insight in \cite{SinghaRoy_20} is the {\em link
  representation} for the entanglement entropies (EE) of all the
bipartitions of a given quantum state. A quantum system consisting of
$N$ parties (qubits or otherwise) will be represented by a link
matrix, $\{J_{ij}\}$, with $i$, $j\in \{1,\cdots,N\}$, with the
property that the entanglement entropy of {\em any subsystem} $A$ can
be approximately computed as a sum of $J$-entries, also called {\em
  link strengths}, associated to all pairs of sites separated by the
partition. In equation form

\beq
S_A \approx \sum_{\substack{i\in A,\\j\in \bar A}} J_{ij},
\label{eq:linkrep}
\eeq
with $J_{ij}\geq 0$. The link matrix may be regarded as an {\em
  adjacency matrix} associated to the entanglement structure:
$J_{ij}>0$ whenever there exists a connection between sites $i$ and
$j$. Thus, the geometry associated to entanglement can be directly
read from the link matrix. In \cite{SinghaRoy_20}, we provided an
algorithm to obtain the optimal entanglement links from the quantum
state. In this work, we will address many relevant questions associated
to the link representation of entanglement, such as the accuracy of
the representation, efficient algorithms to obtain the entanglement
links and their application to the analysis of quantum phases of
matter. 

This work is structured as follows. We provide a summary of known
results in Sec. \ref{sec:theory}. Then, we analyze the accuracy of the
link representation in Sec. \ref{sec:accuracy}. In section
\ref{sec:approx} we will discuss different {\em approximation schemes}
in order to obtain the link matrix. Indeed, the optimal algorithm
shown in \cite{SinghaRoy_20} takes exponential time in $N$. Thus, we
will show more efficient approaches for matrix product states (MPS),
free fermionic states and the case in which contiguous blocks are
specially relevant, among others. Some physical applications to the
analysis of long-range Hamiltonians and the Haldane phase are provided
in Sec \ref{sec:applications}. We conclude with a summary of our main
conclusions and proposals for further work.


\section{Link representation of entanglement}
\label{sec:theory}

This section reviews the main results of Ref. \cite{SinghaRoy_20},
with a slight improvement of the notation and the presentation.

Let us consider a pure state $\ket|\psi>$ of $N$ parties, which will
be qubits in the simplest case. There exist $2^N-2$ possible
subsystems (excluding the block with all sites and no sites), $A$, and
for each of them the entanglement entropy (EE) can be defined as the
von Neumann entropy associated to the reduced density matrix,

\beq
S_A=-\Tr_A(\rho_A \log \rho_A),
\label{eq:vonneumann}
\eeq
with

\beq
\rho_A= \Tr_{\bar{A}} \ket|\psi>\bra<\psi|,
\label{eq:reduced}
\eeq
where $\Tr_A$ stands for the partial trace over subsystem
$A$. Alternatively, different definitions of the EE can be employed,
such as the set of Rényi entropies. Now as for any pure state, $S_A$
obtained for one such subsystem and its compliment yields the same
entropy, the number of subsystem we actually need to consider is
$N_\tot=2^{N-1}-1$.  Let $I\in\{1,\cdots,N_\tot\}$ index the
different subsystems through their binary expansion, and let us
evaluate the EE for all of them,
$\mathcal{S}\equiv\{S_I\}_{I=1}^{N_\tot}$, which we will call the
{\em full entropy data}.

Eq. \eqref{eq:linkrep} can be understood as a linear system of
$N_\tot$ equations (one per $S_I$) to determine $N(N-1)/2$ unknowns,
the link strengths. The system can be formally written as

\begin{figure}[h]
  \includegraphics[width=12cm]{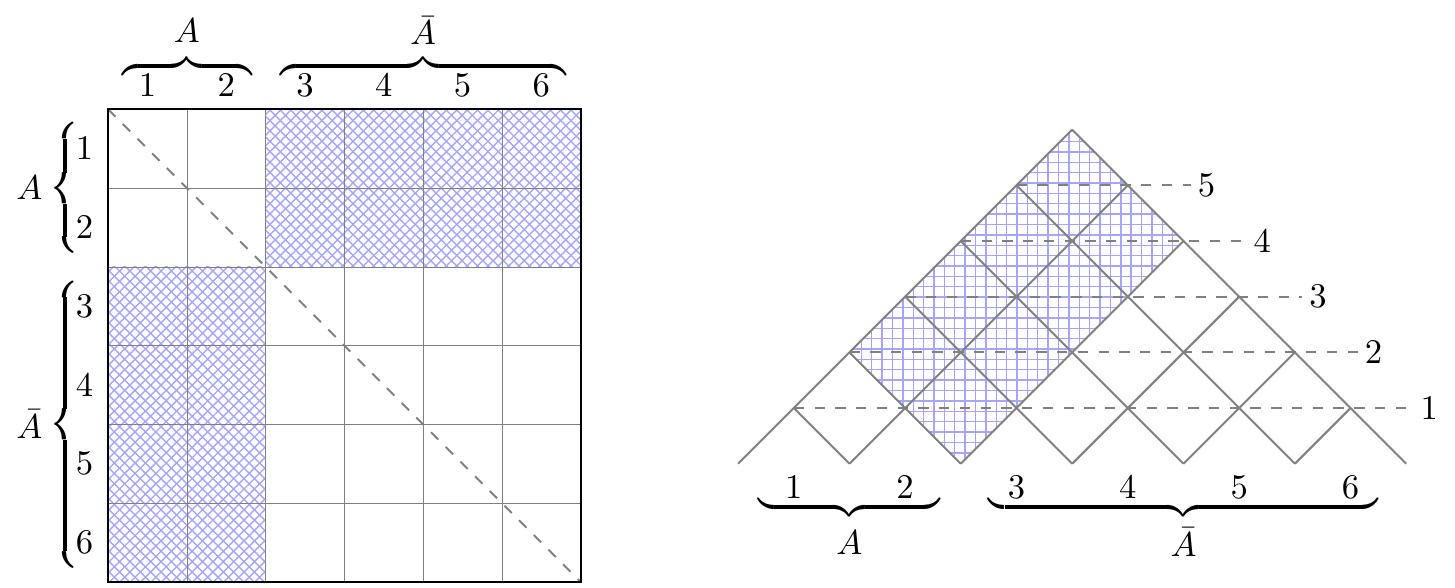}
  \caption{Link representation of the EE associated to the partition
    $A=\{1,2\}$, $\bar A=\{3,4,5,6\}$. Left: link matrix
    representation, the shadowed entries must be added up to build
    $S_A$. Notice the mirror symmetry across the main diagonal. Right:
    triangular view of the link matrix, lying on the main
    diagonal. Horizontal lines correspond to different lengths for the
    associated links. Thus, the first level corresponds to links
    connecting neighboring sites, $J_{1,2}$, $J_{2,3}$, etc. The link
    strengths that must be summed can be obtained in the following
    way. The EE of $A$ or $\bar A$ is obtained adding up all link
    strengths except the right-angled triangles over $A$ and $\bar A$.}
  \label{fig:illust}
\end{figure}

\beq
\mathcal{A}\mathcal{J}=\mathcal{S},
\label{eq:linearsystem}
\eeq
where $\mathcal{J}$ stands for the vector of $N(N-1)/2$ link
strengths, and $\mathcal{A}$ is a matrix with combinatorial origin,
whose entry $\mathcal{A}_{Ik}=1$ iff subsystem $I$ breaks link $k$, or
zero otherwise. Notice that the system \eqref{eq:linearsystem} is
overdetermined, since the number of equations is vastly superior to
the number of unknowns, and in general it has no solutions. Yet,
approximate solutions are still interesting. For example, we can
obtain the optimal solution in the least-squares sense through the
normal equations,

\beq
\mathcal{A}^\dagger\mathcal{A} \mathcal{J} =
\mathcal{A}^\dagger\mathcal{S}.
\label{eq:normaleqs}
\eeq

These normal equations can be efficiently solved, since they are a set
of $N(N-1)/2$ equations with the same number of unknowns, yet the
obtention of $\mathcal{A}^\dagger\mathcal{S}$ requires the knowledge
of the full entropy data. Solving Eq. \eqref{eq:normaleqs} provides
the {\em optimal link representation} for the associated full entropy
data. We denote the $J$-matrix obtained in this way as $J^\opt$ that
we shall use later on.

\begin{figure}
  \includegraphics[width=16cm]{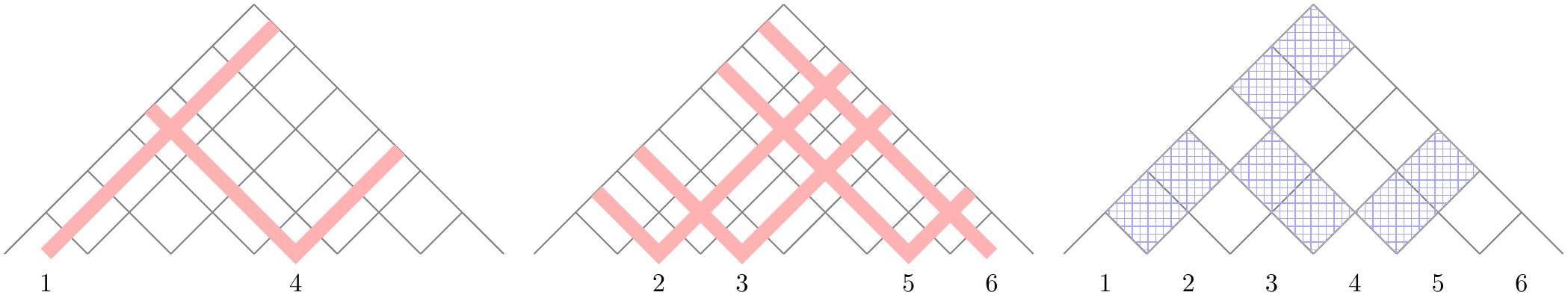}
  \caption{Understanding the triangle view of the link
    representation. In order to find the link strenghts that must be
    added to find the EE of a certain partition we draw the {\em light
      cone} of all sites in $A=\{1,4\}$ (left), then of all sites in
    $\bar A=\{2,3,5,6\}$ (center). The intersection corresponds to the
    link strenghts that contribute to $S_A=S_{\bar A}$.}
  \label{fig:illust_triang}
\end{figure}

\subsection{Entropies from link matrices}

Fig. \ref{fig:illust} provides an illustration to show how to read the
EE of a region following Eq. \eqref{eq:linkrep}. On the left we see
the rectangular view of the link matrix for the block $A=\{1,2\}$,
shading the link strenghts that contribute to $S_A$. On the right we
can see a triangle view of the link matrix, based on the main
diagonal. This representation makes the symmetry $J_{ij}=J_{ji}$
manifest. Moreover, the height of a link strength above the base line
corresponds to the length of the corresponding link.

In Fig. \ref{fig:illust_triang} we provide an alternative route to
obtain the link strengths contributing to the EE of a certain
subsystem $A$. From each site in $A$ we can draw the {\em light cone}
by tracing two diagonal lines at 45 degrees. The light cone of a site
corresponds to the links which start at the given site. The link
strengths to be added are just the intersection of the light cone of
$A$ and that of $\bar A$.

The link representation of the EE has a series of properties which
have been rigorously proved in \cite{SinghaRoy_20}: (a) Symmetry,
$S_A=S_{\bar A}$, which is exact for pure states; (b) Subadditivity:
$S_A+S_B\geq S_{AB}$, which implies the positivity of the mutual
information, $I(A:B)\equiv S_A+S_B-S_{AB}$; (c) Strong subadditivity,
$S_{AB}+S_{BC}\geq S_{ABC}+S_B$.

\subsection{Some exact cases}

The link representation is exact for valence bond states (VBS). For
example, a dimerized state creating bonds between sites $2i$ and
$2i+1$ will have link strengths $J_{2i,2i+1}=\log 2$ and zero
otherwise. The rainbow state, which can be obtained in the strong
inhomogeneity limit from local 1D Hamiltonians is a VBS linking site
$i$ and $N+1-i$, has link strengths $J_{i,N+1-i}=\log 2$ and zero
otherwise. In these cases, the EEs of all partitions are exactly
obtained through the link representation. Conformal states in (1+1)D
have a link representation, in which $J(x,y) = (c/6)/(x-y)^2$ where
$c$ is the central charge and $x, y$ run over the real line. In the
continuum limit we must introduce a short distance regulator
$\epsilon>0$, and define the intervals as
$A_\epsilon=(\epsilon/2,\ell-\epsilon/2)$ and $\bar
A_\epsilon=(-\infty,-\epsilon/2) \cup (\ell+\epsilon/2,\infty)$, thus
obtaining

\beq
S_{A_\epsilon} \simeq
\int_{A_\epsilon} dx \int_{\bar A_\epsilon} dy \frac{ c/6}{ (x-y)^2} =
\frac{c}{3} \log \frac{\ell}{\epsilon}.
\eeq

Indeed, the link strength $J_{ij}$ has an interesting interpretation
from the conformal field theory (CFT) point of view, as the
correlation function of two currents, $J(x,y)=\< \mathbf{J}(x)
\mathbf{J}(y)\>$.

\subsection{Link representation and mutual information}

\begin{figure}
  \includegraphics[width=6cm]{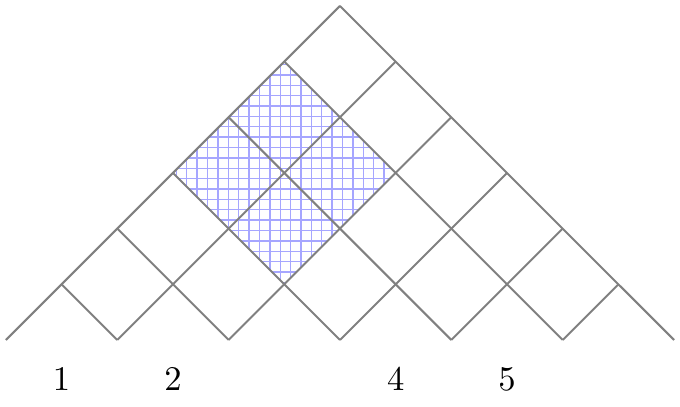}
  \caption{Triangle view of the link representation of the mutual
    information $I(A:B)$ between two intervals, $A=\{1,2\}$ and
    $B=\{4,5\}$, as the diamond spanned by them.}
  \label{fig:mutual}
\end{figure}

Let us remind the definition of mutual information between two
subsystems $A$ and $B$,

\beq
I(A:B)\equiv S_A+S_B-S_{AB}\geq 0,
\label{eq:mutual}
\eeq
which can be obtained easily within the link representation,

\beq
I(A:B)=2\sum_{\substack{i\in A,\\j\in B}} J_{ij}.
\eeq
If $A=\{i\}$ and $B=\{j\}$ we see that $I(i:j)=2J_{ij}$, thus
providing a direct physical interpretation for
$J_{ij}$. Interestingly, the graphical representation of the mutual
information between two subsystems is simpler than for the EE, as it
can be checked in Fig. \ref{fig:mutual}.

\subsection{Entanglement contour}

An {\em entanglement contour} associated to a given subsystem is a
partition of the EE among its sites \cite{Vidal_14} (see also \cite{BPE1, BPE2}), $s_A(i)\geq 0$,
such that $\sum_i s_A(i)=S_A$. The link representation provides an
entanglement contour for each possible subsystem,

\beq
s_A(i)=\sum_{j\in \bar A} J_{ij}.
\label{eq:contour}
\eeq

Notice that, in similarity to the link representation, the
entanglement contour is not uniquely defined. Indeed, the contour for
a single block $A$ is {\em underdetermined}, i.e. there are many
possible contours $\{s_A(i)\}_{i=1}^{|A|}$ yielding the same EE
$S_A$. Yet, physically motivated definitions tend to coincide
\cite{Tonni_18,Alba_19}.


\section{Accuracy of the link representation}
\label{sec:accuracy}

The accuracy of the optimal link representation, given by the solution
of Eq. \eqref{eq:normaleqs}, can be quantified as follows. Let us
obtain the EE of all blocks within the link representation, $\{\hat
S_I\}_{I=1}^{N_\tot}$, and then compare these values to the exact
ones, $\{S_I\}_{I=1}^{N_\tot}$, obtaining the average absolute
error,

\beq
\Delta S \equiv {1\over N_\tot} \sum_I |\hat S_I-S_I|.
\label{eq:abserror}
\eeq
This average error can be properly normalized dividing each partition
error by its entropy, giving rise to the average relative error,

\beq
\Delta_R S \equiv {1\over N_\tot} \sum_I {|\hat S_I-S_I| \over S_I}.
\label{eq:exprelerror}
\eeq
Yet, this definition is not very convenient, because some EE can be
exactly zero. Thus, it is more relevant to define the {\em average EE}
over all partitions, $\<S\>$, and use it to define a relative error

\beq
\varepsilon = {\Delta S \over \<S\>}.
\label{eq:error}
\eeq

On occasions, it is also convenient to restrict the error measure over
some subset of partition. In this case, we will use a suitable
notation to highlight that we are not extending the measure to the
full entropy data.

As a benchmark, we will employ two reference states: the ground state (GS) of a
conformally invariant free-fermionic state, and a random state.

\subsection{Free fermionic states}

Let us consider the Hamiltonian

\beq
H=-\sum_{i} t_{i,i+1} c^\dagger_i c_{i+1} + \text{h.c.},
\label{eq:freefermions}
\eeq
where $c_i$ is the fermionic annihilation operator on site $i$ with
either open boundary conditions (OBC) or periodic (PBC), in this case
using only $N=2$ mod 4 to avoid degeneracies. The GS of this
Hamiltonian is a Slater determinant, which can be written as
$\ket|\Psi>=\prod_{k=1}^{N/2} b^\dagger_k \ket|0>$, with
$b^\dagger_k=\sum_i U_{k,i} c^\dagger_i$ and $U$ is the matrix
diagonalizing the hopping matrix $UTU^\dagger=\tilde{T}$, with
$T_{i,i+1}=T_{i+1,i}=t_{i,i+1}$. We should stress that the free
fermionic GS with $t_{i,i+1}=1$ does not follow a strict area
law. Instead, the entanglement entropy of a contiguous block of size
$\ell$ grows like $S(\ell)\approx (c/3)\log\ell$, due to conformal
invariance. We will also consider a dimerized Hamiltonian,

\beq
H=-\sum_{i} (1+(-1)^i\delta)c^\dagger_i c_{i+1} + \text{h.c.},
\label{eq:dimerized}
\eeq
which is not conformally invariant, and presents a mass gap.

\begin{figure}
  \includegraphics[width=13cm]{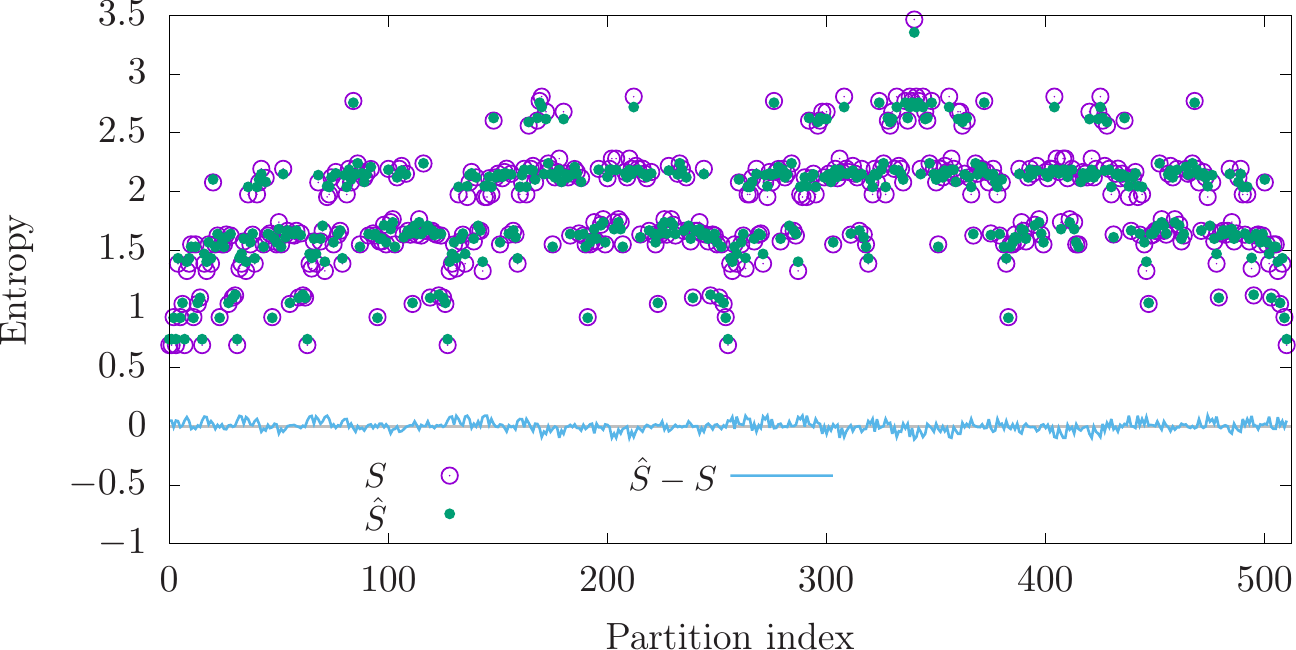}
  \caption{Entropies of all subsystems, in lexicographical order,
    along with the (signed) absolute errors associated to the optimal
    link representation for  free fermion state with $N=10$ sites. Here the empty circles correspond to the exact values of the entropies, $S_I$, while the full dots correspond to the optimal values obtained using the link representation, $\hat S_I$.}
  \label{fig:error_ff}
\end{figure}

Fig. \ref{fig:error_ff} shows the entropies for all partitions of the
free fermion state with $t_{i,i+1}=t$ and $N=10$ sites. Actually, not
all the subsystems are shown, because due to the symmetry
$A\leftrightarrow \bar{A}$, we may disregard those with the last site
absent, thus retaining only $N_\tot=2^{N-1}-1$. The empty circles
correspond to the exact values of the entropies, $S_I$, while the full
dots correspond to the optimal values obtained using the link
representation, $\hat S_I$. The continuous line below shows the signed
absolute error, $\hat S_I-S_I$, which averages to zero.

\begin{figure}
  \includegraphics[width=8cm]{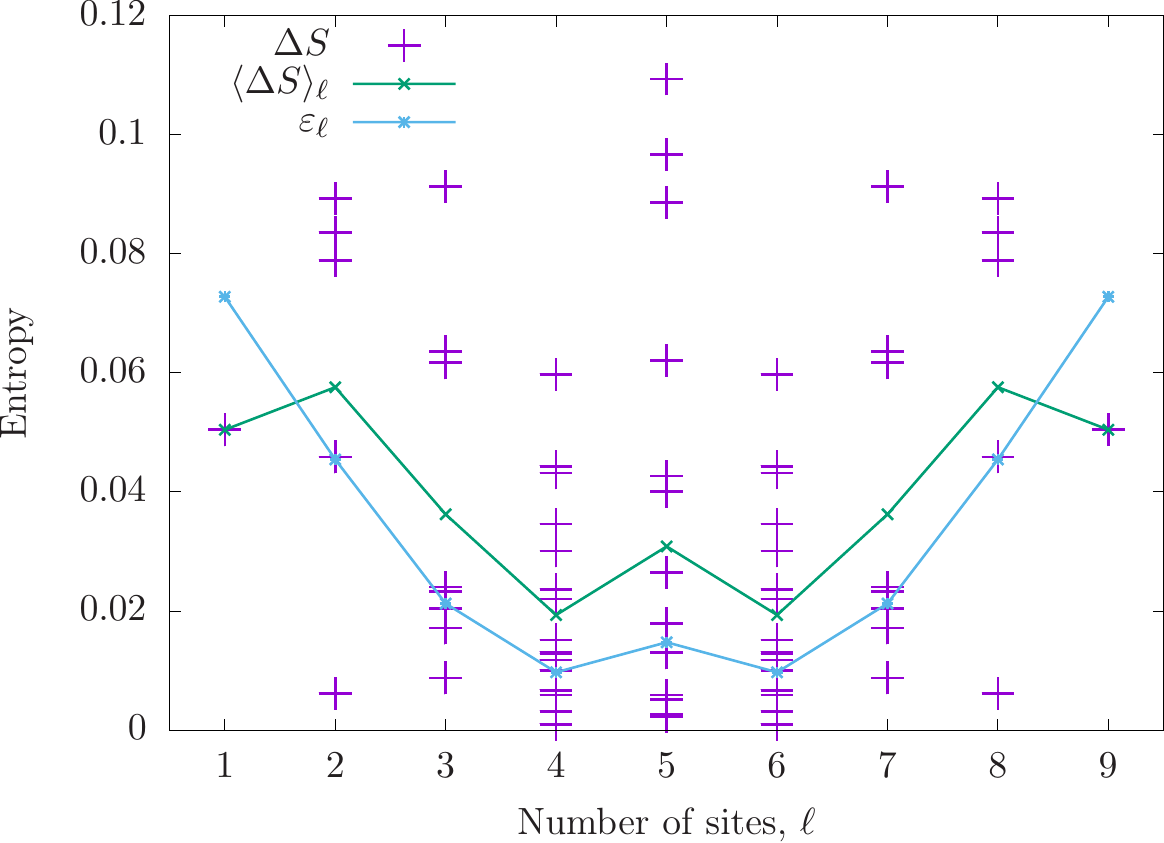}
  \includegraphics[width=8cm]{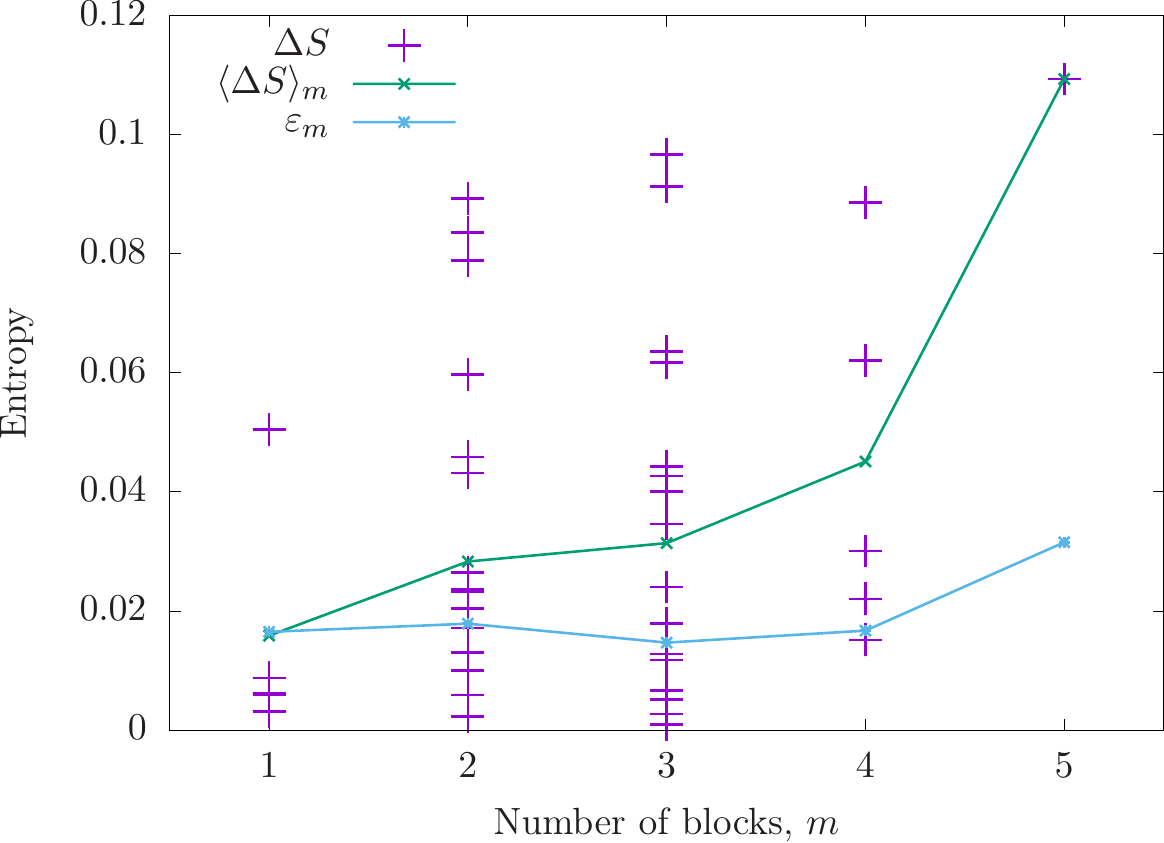}
  \caption{Accuracy of the link representation for the free fermionic
    state. Left: the purple crosses represent the entropy absolute
    errors classified by the number of sites in our block, the green
    line denotes the average of this absolute error for each block
    size, and the cyan line divides this value by the average entropy
    corresponding to each block size, yielding a relative
    error. Right: same data, classified by the number of blocks in
    each partition.}
  \label{fig:struct_ff}
\end{figure}

Notice that absolute errors shown in Fig. \ref{fig:error_ff} are very
tiny. Indeed, the average relative error, as defined in
Eq. \eqref{eq:error}, is 1.7\%. Are there any systematic trends for
different types of blocks? Indeed, Fig. \ref{fig:error_ff} presents
noticeable patterns. Some blocks of neighboring entropies present
negligible error, while other groups present larger
deviations. Fig. \ref{fig:struct_ff} clusters the data as a function
of the number of sites (left) and the number of blocks (right) in each
partition. Regarding the dependence of the error on the number of
sites, $\ell$ we observe that the natural symmetry
$\ell\leftrightarrow N-\ell$ is respected. The relative error is
lowest for number of sites close to $N/2$. The reason is that the
optimal link representation attempts to fit all entropy data
simultaneously, and the number of partitions with $\ell\sim N/2$ is
higher than for either small or large sizes. On the other hand, we
observe that the relative error is approximately constant for all
block sizes, even though the absolute error grows with the number of
blocks.

\subsection{Size dependence}

\begin{figure}
  \includegraphics[width=10cm]{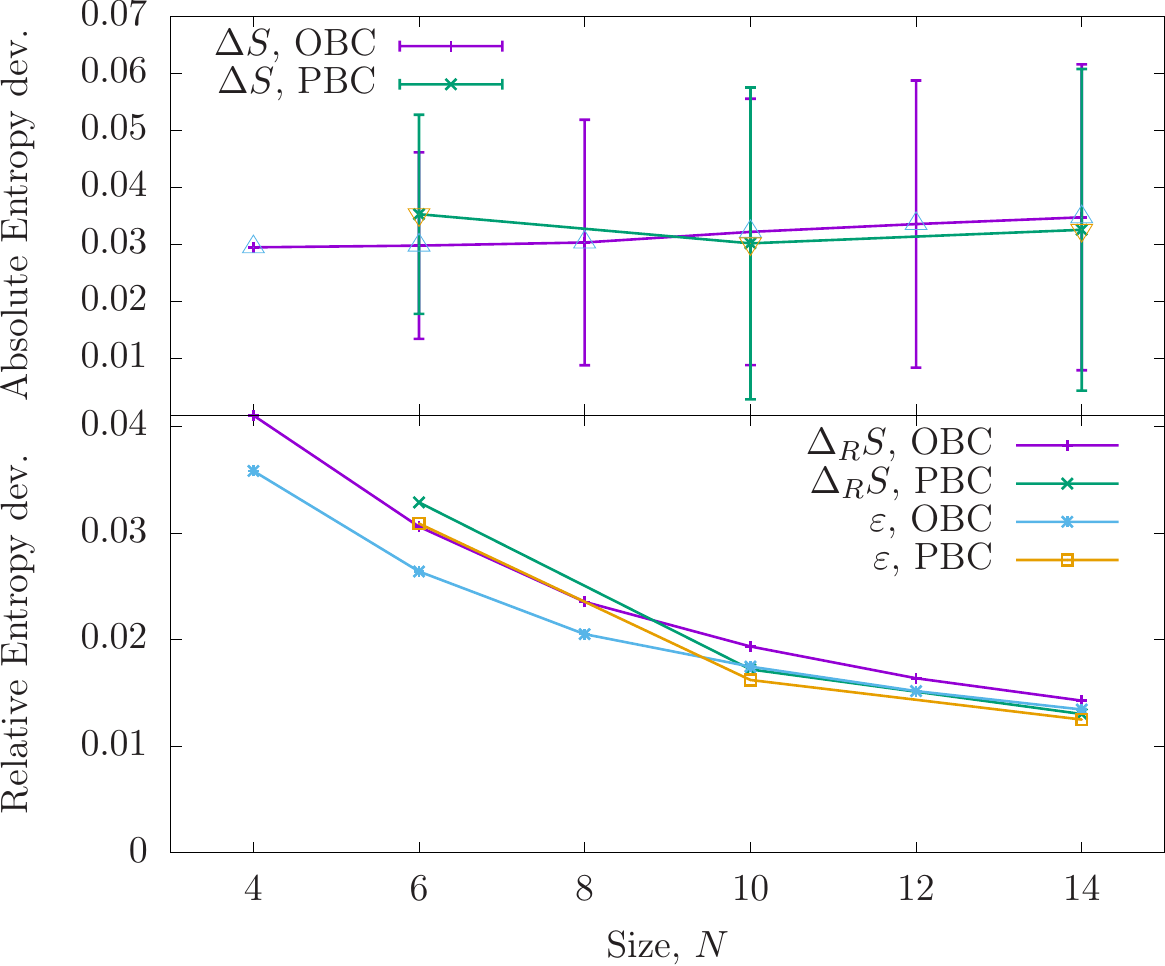}
  \caption{Top: Absolute mismatch between the exact and the approximate
    entropies of the GS of fermionic chains with OBC and PBC, along
    with their deviations, for different system sizes. Bottom:
    Relative mismatch between the exact and approximate entropies,
    along with the error defined in Eq. \eqref{eq:error}.}
  \label{fig:error_N}
\end{figure}

It is important to ask how does the accuracy of the link
representation scale with the system size. Unfortunately,
Eq. \eqref{eq:normaleqs} can not be solved for large system sizes. We
have obtained the absolute and relative deviations,
Eq. \eqref{eq:abserror} and \eqref{eq:exprelerror} along with the {\em
  error} defined through Eq. \eqref{eq:error} for the GS of open and
periodic fermionic chains up to $N=14$. In the periodic case, we have
restricted the computation to values of $N=2$ mod 4, because only in
that case the GS is unique. The results are shown in
Fig. \ref{fig:error_N}.

We observe that though the absolute error remains almost constant, the
relative errors decay as a function of the system size, thus showing
that the optimal link representation gets more and more accurate as
the system size increases. These results suggest that the optimal link
representation may become exact in the thermodynamic limit for these
conformally invariant states. Unfortunately, the huge computational
effort required to obtain the optimal link representation forces us to
attempt alternative approximations.

\subsection{Random states}

Let us consider random pure states of $N$ qubits under a Haar
measure. They can be sampled by choosing the real and imaginary parts
of each component on any basis from a standard Gaussian distribution,
and normalizing afterwards. The entanglement entropy of these states
has been known for a long time \cite{Page_93}. Indeed, it was shown
that the average entropy of a subsystem only depends on the minimum
between the number of sites of the aforementioned subsystem and those
of the complementary,

\beq
\<S(\ell)\> \approx \ell  \log 2 - {1\over 2^{N-2\ell+1}}.
\label{eq:page}
\eeq

\begin{figure}
  \includegraphics[width=13cm]{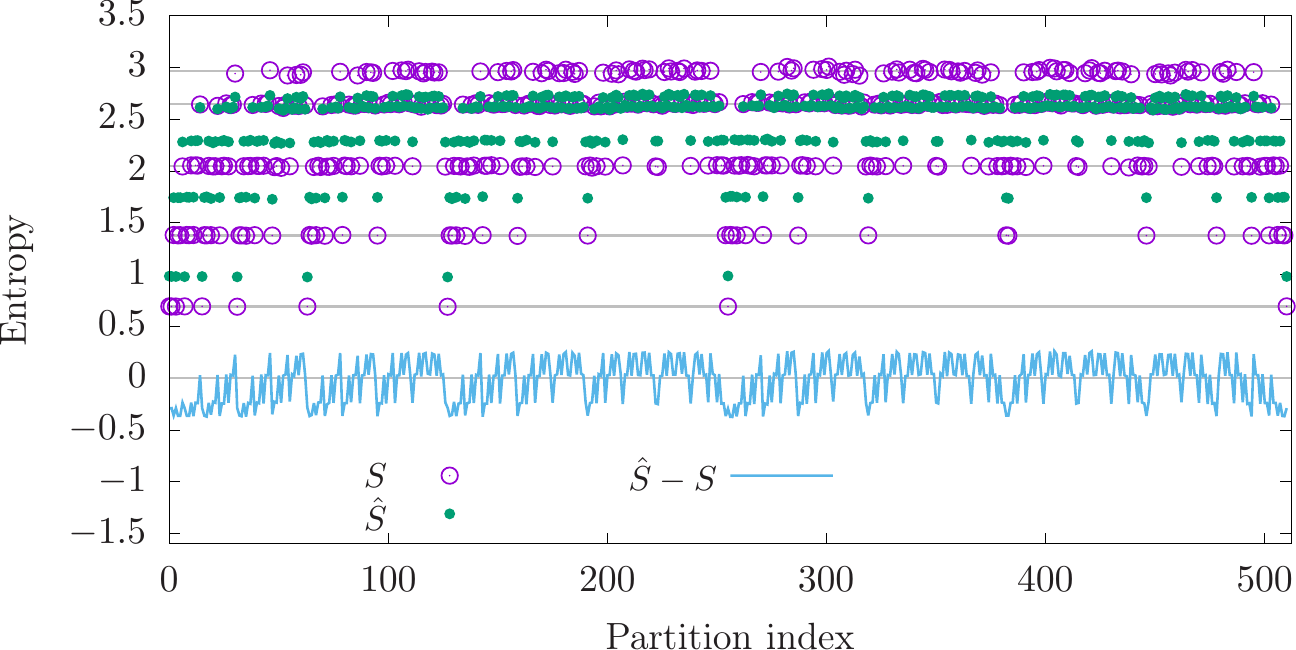}
  \caption{Entropies of all subsystems, in lexicographical order,
    along with the (signed) absolute errors associated to the optimal
    link representation for a realization of the random state defined
    in the text. Here the empty circles correspond to the exact values of the entropies, $S_I$, while the full dots correspond to the optimal values obtained using the link representation, $\hat S_I$. The horizontal bars mark the expectation values for
    the entropies according to Page's law for each block size,
    following Eq. \eqref{eq:page}.}
  \label{fig:error_random}
\end{figure}

Naturally, these averages are taken for many states and
subsystems. Yet, a single realization shows the remarkable accuracy of
Eq. \eqref{eq:page}, as we can see in Fig. \ref{fig:error_random},
where all the entropies of a single random state have been depicted in
empty circles, as in Fig. \ref{fig:error_ff}, with the approximation
according to the optimal link representation in full circles. The
horizontal grey lines correspond to the theoretical predictions for
the average of the entropy for each block size,
Eq. \eqref{eq:page}. We notice that the exact values fluctuate weakly
around the theoretical predictions, with larger fluctuations
corresponding to larger sizes. 

\begin{figure}
  \includegraphics[width=8cm]{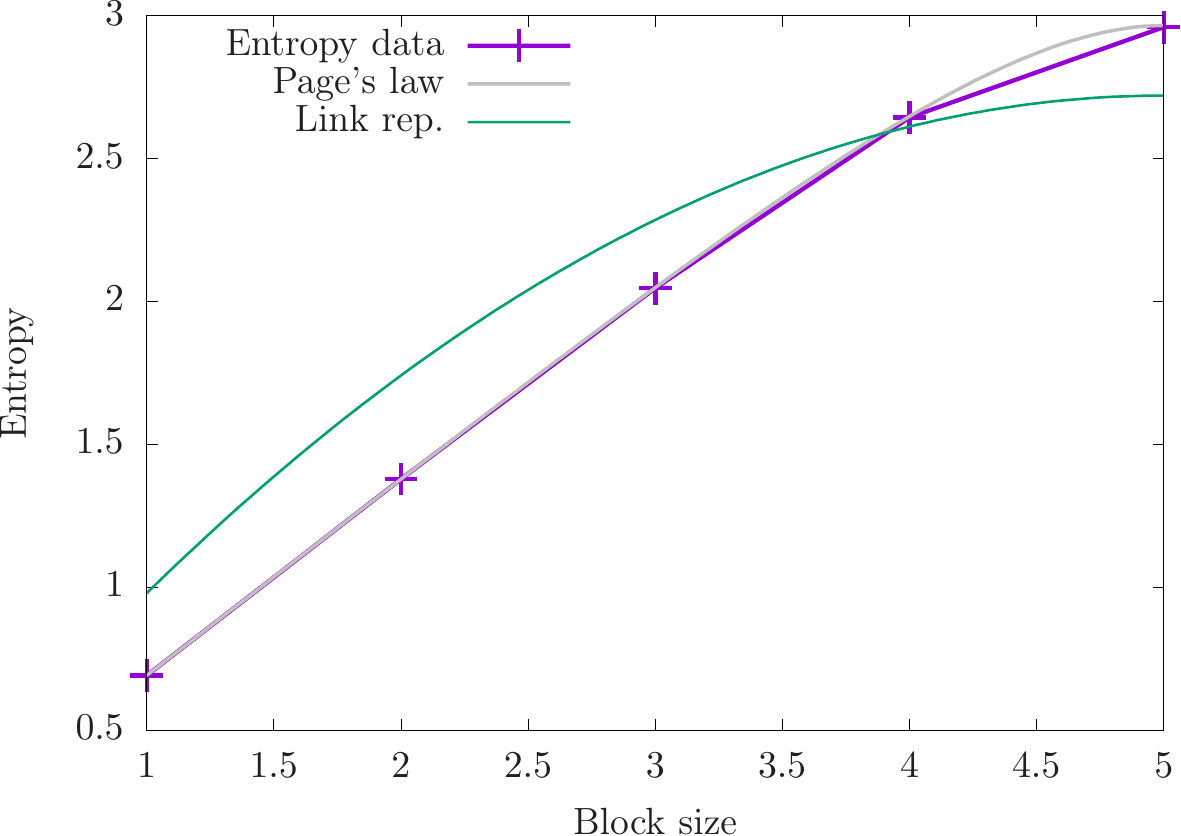}
  \caption{Comparison of the predictions of Page's law and the optimal
    link representation for the average entropies of a random
    state.}
  \label{fig:struct_random}
\end{figure}

The relative errors are much higher than in the free fermionic case,
around 7\%. The approximate values obtained with the optimal link
representation also appear in horizontal lines corresponding to block
sizes, but these lines do not correspond to the exact ones. The
effective permutation invariance of the state gives rise to nearly
constant link strengths, $J_{ij}\approx \chi$ for all $i$, $j$. In our
case, we obtain $\chi=0.157\pm 0.004$. The prediction for the entropy
of a block of size $\ell$, according to the optimal link
representation is

\beq
S(\ell)=\chi\;\ell(N-\ell).
\label{eq:random_state_lr}
\eeq
Obtaining the link representation for these states amounts to an
attempt to fit Page's law using Eq. \eqref{eq:random_state_lr}, as we
can see in Fig. \ref{fig:struct_random}. For $\ell\ll N$, a better
link representation is obtained by choosing $\chi=N^{-1}$. 

Indeed, the link representation is not specially good for random
states. The reason is that it has been designed for states for which
{\em an area law} may emerge, either through the geometry of the
Hamiltonian or a different geometry, associated to entanglement.


\section{Approximation methods for the entanglement links}
\label{sec:approx}

Obtaining the optimal link representation associated to a pure state
is a hard computational problem, in general terms, since it requires
the determination of the full entropy data, in principle $2^N$
entropies, although it reduces in practice to $N_\tot=2^{N-1}-1$
independent values. Obtaining the EE of arbitrary partitions for
arbitrary quantum states is a costly procedure itself. Thus, it is
specially relevant to devise efficient approximation
techniques. Ideally, that requires a polynomial amount of resources.

Finding an efficient approximation for the link matrix is not only a
technical necessity. It is also a problem of fundamental relevance:
can we obtain a link representation for GS of local Hamiltonians?

\begin{figure}
  \includegraphics[width=10cm]{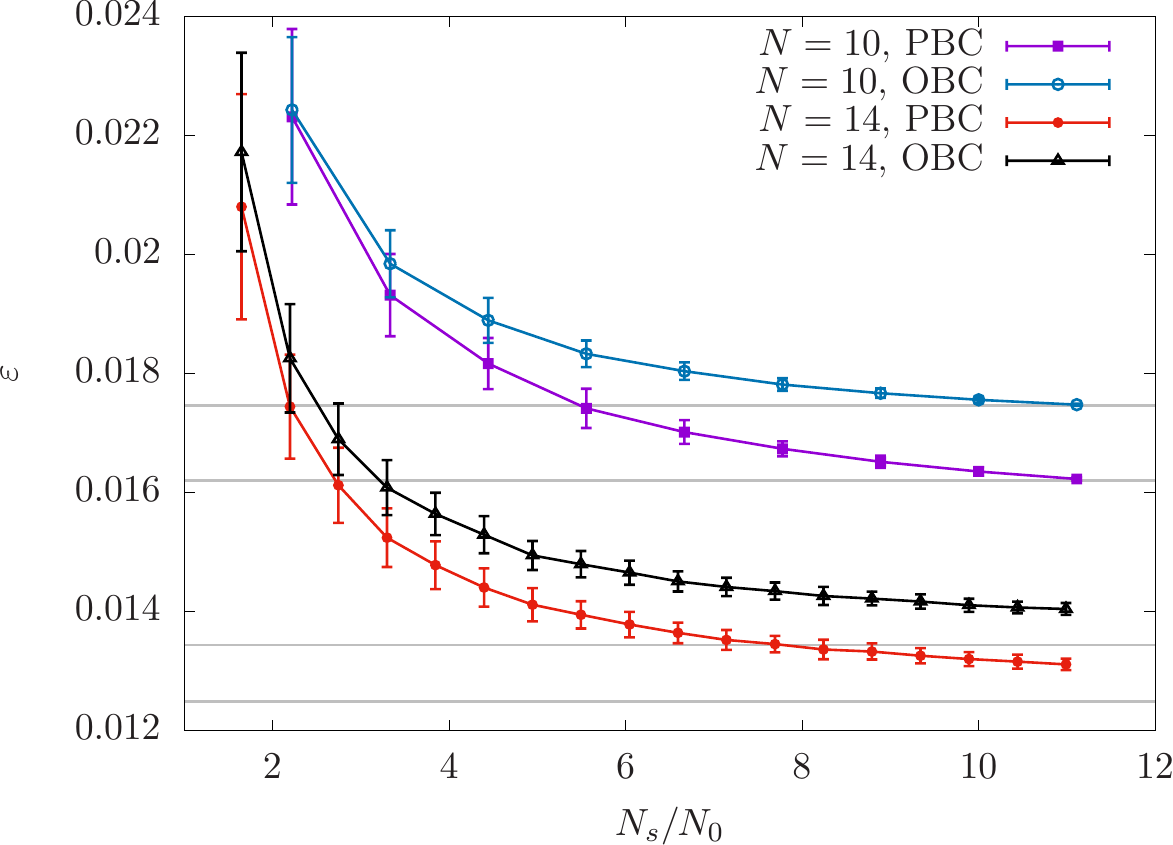}
  \caption{Error of the link representation obtained with $N_s$
    samples of the partition space, scaled with $N_0=N(N-1)/2$, for
    four different fermionic GS, with OBC and PBC, $N=10$ and 14. Gray
  lines correspond to the error associated to the optimal link
  representation, see Fig. \ref{fig:error_N}.}
  \label{fig:sampling}
\end{figure}

\subsection{Random sampling}

A first approximation scheme can be built renouncing to consider an
exponential number of partition, and {\em sampling} a number $N_s$ of
them instead. The results are indeed promising. Of course, $N_s$ must
be larger than $N(N-1)/2\equiv N_0$, which is the number of
independent link strengths. Fig. \ref{fig:sampling} shows the average
error associated to the link representation obtained sampling $N_s$
partitions, as a function of $N_s/N_0$ for two fermionic GS, with open
and periodic boundaries (OBC and PBC) and two sizes, $N=10$ and
$N=14$, with the associated error bars. The horizontal lines
correspond to the errors associated to the optimal link
representation. Each experiment was repeated 100 times in order to
ensure reproducibility of our results.

From Fig. \ref{fig:sampling} we are led to conjecture that for a state
with a good link representation, sampling $N_s\approx 10\times N_0$
should be enough for reasonably accurate values of the link
strengths. This is indeed good news, since it means that we should
obtain the entropies of $O(N^2)$ partitions and solve a linear system
of a similar size, thus rendering the problem tractable.

\subsection{Entanglement links for contiguous blocks}

Let us consider the possibility that we do not have access to the full
entropy data, but only the EE corresponding to contiguous blocks on a
length $N$ chain with PBC. A proper notation can be rather
helpful. Let us define

\beq
A_{i,j}=\{i,\cdots,j-1\},
\label{eq:notA}
\eeq
where site indices are always considered mod $N$, and
$S_{i,j}=S[A_{i,j}]$ is the associated EE. Thus, $S_{i,i}=0$ for any
pure state, because $A_{i,i}$ is always the whole chain. Moreover,
$S_{i,i+1}$ stands for the EE of block $A_{i,i+1}=\{i\}$. Naturally,
for a pure state we have $S_{i,j}=S_{j,i}$, because both blocks are
complementary, and this symmetry property motivates the definition.

The entropies $S_{i,j}$ can be expressed through

\beq
S_{i,j}=\sum_{k=i}^{j-1}\sum_{l=j}^{i-1} J_{k,l},
\label{eq:Sij}
\eeq
where the summations must be understood mod $N$. Let us obtain its
difference in the first index,

\beq
(\Delta_1 S)_{i,j}\equiv S_{i+1,j}-S_{i,j}=
\(\sum_{k=i+1}^{j-1} - \sum_{k=j}^i\) J_{i,k},
\label{eq:DS}
\eeq
and now, let us obtain its difference in the second index,

\begin{align}
(\Delta_2 \Delta_1 S)_{i,j}=& (\Delta_1 S)_{i,j+1} - (\Delta_1 S)_{i,j}
\nonumber\\
=&\(\sum_{k=i+1}^{j} - \sum_{k=j+1}^i 
- \sum_{k=i+1}^{j-1} + \sum_{k=j}^i\) J_{i,k} \nonumber\\
=& S_{i+1,j+1}-S_{i+1,j}-S_{i,j+1}+S_{i,j}\nonumber\\
=& 2 J_{i,j}.
\label{eq:D2S}
\end{align}

The continuous version of this equation is simply

\beq
J(x,y)={1\over 2} {\partial^2 S(x,y)\over \partial x \partial y},
\label{eq:JD2S}
\eeq
where we have assumed that $S(x,y)$ is the EE of the interval
$[x,y]$. The validity of Eq. \eqref{eq:D2S} can be checked graphically
in Fig. \ref{fig:d2s} in Appendix\ref{AppendixA}.

Notice that the equation of
$2J_{i,i+1}=S_{i+1,i+2}-S_{i+1,i+1}-S_{i,i+2}+S_{i,i+1}=S[i]+S[i+1]-S[i,i+1]$,
i.e. it is the mutual information between the neighboring sites. For
longer links this equation is not valid. Moreover, in the case of a
translation invariant state Eq. \eqref{eq:D2S} reduces to

\beq
J_{i,i+r}=S(r)-{1\over 2}S(r-1)-{1\over 2}S(r+1),
\label{eq:blockTI}
\eeq
where $S(r)$ denotes the EE of all contiguous blocks of size $r$. In
the continuous limit, Eq. \eqref{eq:blockTI} becomes

\beq
J(\ell)=-{1\over 2} {\partial^2 S(\ell) \over \partial \ell^2}.
\label{eq:Jell}
\eeq
These two last equations were already shown in
Ref. \cite{SinghaRoy_20}.

\bigskip

We may generalize the above expression in order to build an
approximation to the entanglement links which only considers
partitions containing $n_B$ contiguous blocks or less. Thus, the
aforementioned contiguous block approximation would correspond to the
$n_B=1$ case. Naturally, the quality of this approximation will grow
with $n_B$ up to the maximal value, $n_B=N/2$. Table
\ref{table:structured} shows the relative errors associated to this
structured approximation for a free-fermionic GS with PBC.

\begin{table}
\begin{tabular}{l|c|c|c}
  & \multicolumn{3}{c}{$\varepsilon$} \\
  \hline
  $n_B$ &  $N=6$ & $N=10$ & $N=14$ \\
  \hline
  1 & 0.0611 & 0.0663  &  0.0681 \\  
  2 &  0.0272 & 0.0281&  0.0362 \\
  3 & 0.0308 & 0.0173  &  0.0187 \\
  4 &        &  0.0161  &   0.0133 \\
  5 &        & 0.0162  &   0.0125\\
  6 &        &         &  0.0124\\
  7 &        &         &  0.0124\\
\end{tabular}
\caption{Relative errors within the structured approximation, as a
  function of the number of blocks considered in the partition for a free-fermionic GS with PBC.}
\label{table:structured}
\end{table}

\subsection{Approximation for free-fermionic states}

Let us consider a Gaussian fermionic state, written in the form of a
Slater determinant, $\ket|\Psi>=\prod_k b^\dagger_k \ket|0>$, with
$b^\dagger_k=\sum_i U_{k,i} c^\dagger_i$, where $c^\dagger_i$ is the
creation operator on the $i$-th site. The entanglement properties
associated to a block $A$ can be obtained through the correlation
submatrix, $(C_A)_{i,j}=\sum_k \bar U_{k,i} U_{k,j}$, with $i$, $j\in
A$. Let $C_A=W_A\Lambda_AW^\dagger_A$, where
$\Lambda_A=\text{diag}(\nu^A_1,\cdots,\nu^A_{|A|})$. Then,

\beq
S_A=\sum_{p=1}^{|A|} H_2(\nu_p),
\label{eq:SA}
\eeq
where $H_2(x)\equiv -x\log(x)-(1-x)\log(1-x)$. Interestingly, the
eigenvectors of $C_A$, given in the columns of $W_A$, are usually
disregarded, except in the evaluation of the {\em entanglement
  contour}, which provides a possible partition of the EE among the
different sites of the block,

\beq
s_A(i)=\sum_{p=1}^{|A|} H_2(\nu_p) |(W_A)_{p,i}|^2,
\label{eq:contour}
\eeq
from which it is clear that $s_A(i)\geq 0$ and $\sum_{i\in
  A}s_A(i)=S_A$. Let us also consider the complementary block, $\bar
A$. Indeed, $\Lambda_A$ and $\Lambda_{\bar A}$ are easily related: for
every eigenvalue $\nu^A_p\in (0,1)$ there must be corresponding
eigenvalue $\nu^{\bar A}_{q(p)}=1-\nu^A_p$. Barring degeneracies, each
eigenvector $(W_A)_p$ of $C_A$ corresponds to an eigenvector of
$(W_{\bar A})_{q(p)}$ of $C_{\bar A}$, and we may write the
approximation

\beq
J_{i,j}\simeq\sum_{p=1}^{|A|} H_2(\nu^A_p)
|(W_A)_{p,i}(W_{\bar A})_{q(p),j}|^2.
\label{eq:Jslater}
\eeq
In similarity with the equation for the entanglement contour,
Eq. \eqref{eq:contour}, Eq. \eqref{eq:Jslater} obtains the
entanglement links by partitioning the entanglement entropy into
contributions stemming from pairs of sites. In practice,
Eq. \eqref{eq:Jslater} presents several drawbacks. First, the
entanglement links obtained correspond to a single partition. A
possible solution is to consider a finite set of partitions and
average the values of the entanglement links over them. Moreover, the
spectrum $\{\nu^A_p\}$ is very often exactly degenerate, thus
preventing us from a clear association between the eigenvalues of $A$
and $\bar A$.

These drawbacks lead us to propose a concrete application for the
simplest case, where we consider the set of partitions containing a
single site, $A_i=\{i\}$. In this case, the spectrum of $C_{A_i}$
contains a single entry, $C_{i,i}$, and the spectrum of $C_{\bar A_i}$
contains a single non-trivial entry, $1-C_{i,i}$. Thus, we can write

\beq
J_{i,j}={1\over 2} \(H_2(C_{i,i}) |B_{i,j}|^2+H_2(C_{j,j})|B_{j,i}|^2\),
\label{eq:approxslater}
\eeq
where $B_{i,j}=(W_{\bar A_i})_{q,j}$ is the $j$-th component of the
eigenvector associated to the single non-trivial eigenvalue of
$C_{\bar A_i}$. The advantage of this approach is its extreme
simplicity and low computational cost.

\begin{figure*}
  \includegraphics[width=8cm]{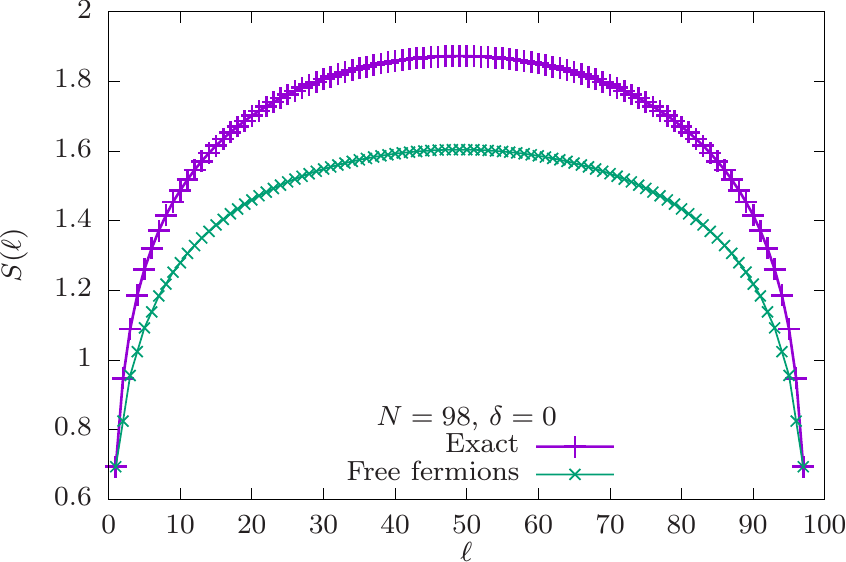}
  \includegraphics[width=7cm]{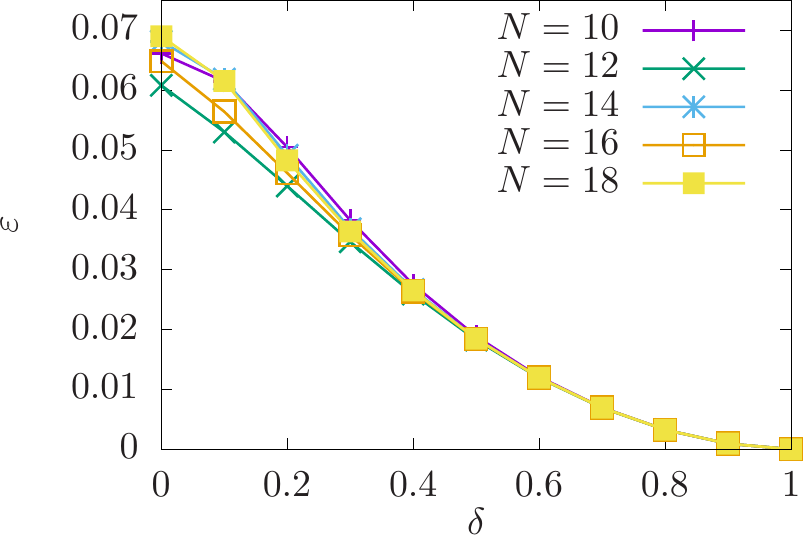}
  \includegraphics[width=10cm]{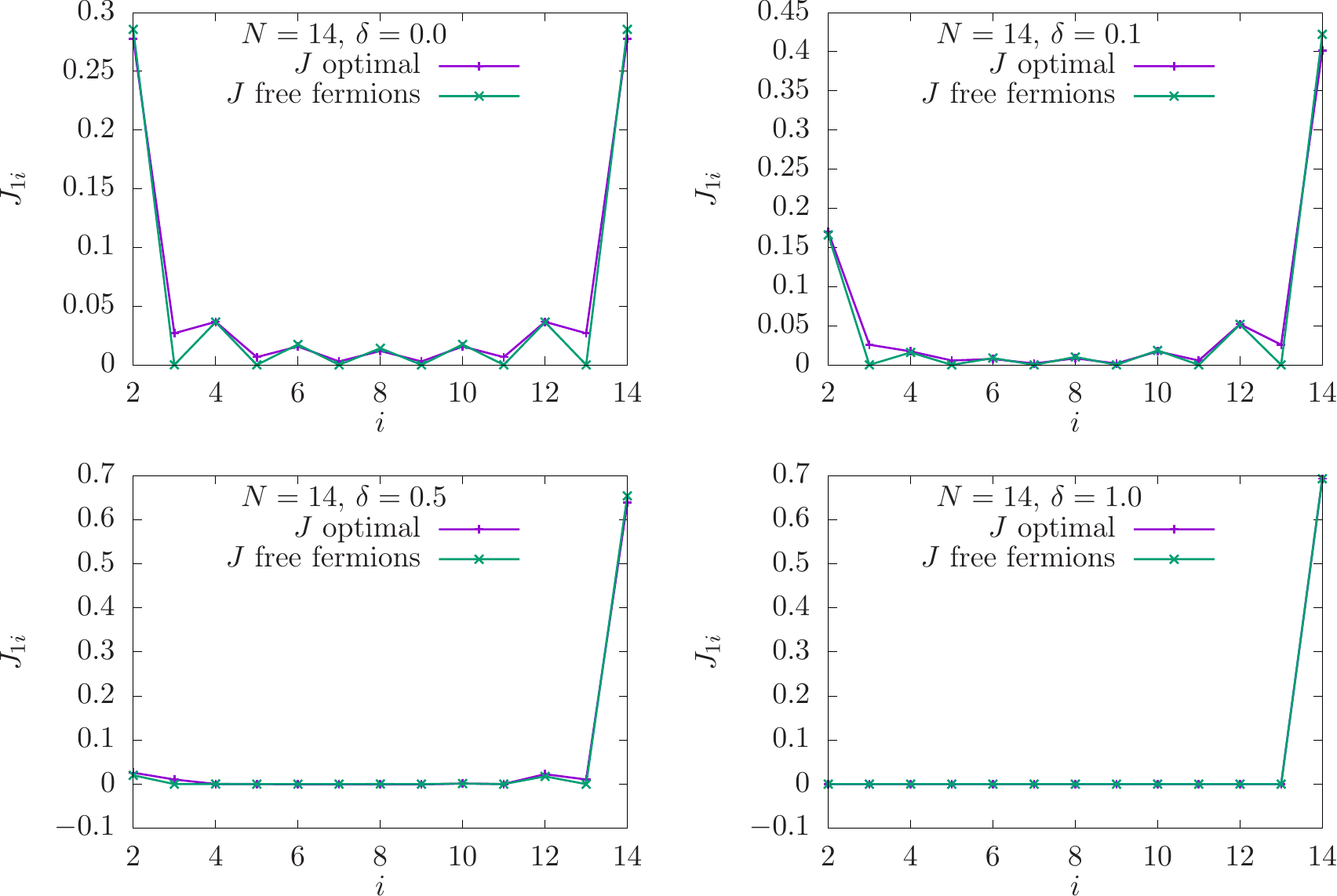}
  \caption{(a) Entropies $S[1,\cdots,\ell]$ as a function of $\ell$
    for the same system, comparing with the compact-block
    approximation and the free-fermionic approximation for a chain
    with $N=98$ sites, which is clearly not accurate. (b) Error of the
    free-fermionic approximation as a function of $\delta$ for several
    sizes. Notice that they are approximately equal, and decay to zero
    as $\delta$ increases. (c) Values of $J_{i,i+n}$ for a dimerized
    free-fermionic chain as a function of $n$ for several values of
    $\delta$. Notice that the even values are correctly represented,
    but not the odd ones, which are zero in the free-fermionic
    approximation.}
  \label{fig:slater}
\end{figure*}

We should remark that this approximation is exact for partitions
containing a single site, but for other blocks the accuracy can be
much less than that obtained for the optimal link matrix. Indeed, as
we shall show, the EE of large contiguous blocks can be subject to
larger errors, while the global accuracy is still good enough because
of the large number of sparse partitions.

The top-left panel of Fig. \ref{fig:slater} shows the EE of a
contiguous block starting from the left for a periodic fermionic chain
with $N=98$, obtained within the contiguous blocks approximation
(which is exact in this case by construction) and for the
free-fermionic approximation. Indeed, the error for those EE is
large. In order to understand why we may take a look at the bottom
panel of Fig. \ref{fig:slater}, where we see the optimal link
strengths compared to the free-fermion approximation for different
distances, $J_{i,i+n}$. The most striking difference is the fact that
the exact link strengths for odd $n$ are non-zero, but the
free-fermionic approximation is zero. The non-zero values of
$J_{i,i+n}$ are very accurately obtained with the free-fermionic
approximation. Thus, we are led to the conjecture that the
free-fermionic approximation imposes a structure on the link
representation which is not optimal, related to the parity
oscillations associated to the Fermi momentum $k_F=\pi/2$.

The top-right panel of Fig. \ref{fig:slater} provides some good news
regarding the relative error associated to the free-fermionic link
representation for different sizes of the dimerized GS,
Eq. \eqref{eq:dimerized}, as a function of $\delta$. We notice that
the error is a few percent for $\delta\to 0$, several times larger
than the optimal value, and nearly independent of the system size
$N$. As $\delta$ increases, the error is reduced, and for large
$\delta$ we approach a valence bond state for which the error
vanishes. Since this representation is extremely cheap to obtain, in
terms of computational resources, it seems to be a good choice only
for states which are close to a VBS.

\subsection{Approximation for matrix product states}

Matrix product states provide the main examples for states fulfilling
the area law of entanglement. Therefore, we expect them to have a
specially simple link representation, with exponentially decaying link
strengths, $J_{i,j}\approx \exp(-|i-j|/\xi)$, for some correlation
length $\xi$ \cite{Chen}. Indeed, this is our main result, but the technical
details are also relevant. We will provide explicit expressions for
the link strengths based on two different approximations. First, as we
mentioned before, the link strengths can be approximated as the mutual
information between pairs of sites. Then, we will approximate the link
strengths from the entropies of contiguous blocks, via
Eq. \eqref{eq:blockTI}.

\bigskip

\subsubsection{Mutual information approximation}

Let us consider a translational invariant matrix product state (MPS)
representation of a general quantum many-body state $\ket|\Psi>$ with
periodic boundary condition \cite{Verstraete_08,Cirac_09,Orus_14},
given by

\begin{eqnarray}
\ket|\Psi>&=&\sum_{i_1 i_2 \dots i_N} \text{Tr}(A_{i_1} A_{i_2}\dots
A_{i_N}) \ket|i_1 i_2\dots i_N>,
\label{MPS_general}
\end{eqnarray}
where $A_{i_k}$ are $D\times D$ matrix. Now starting from the above
expression our aim is to find the reduced density matrix corresponding
to the block formed by sites $i$ and $j$, situated at a distance
$l+1=|j-i|$ from each other. In other words, we wish to compute

\begin{figure}[h]
\includegraphics[width=6.5cm]{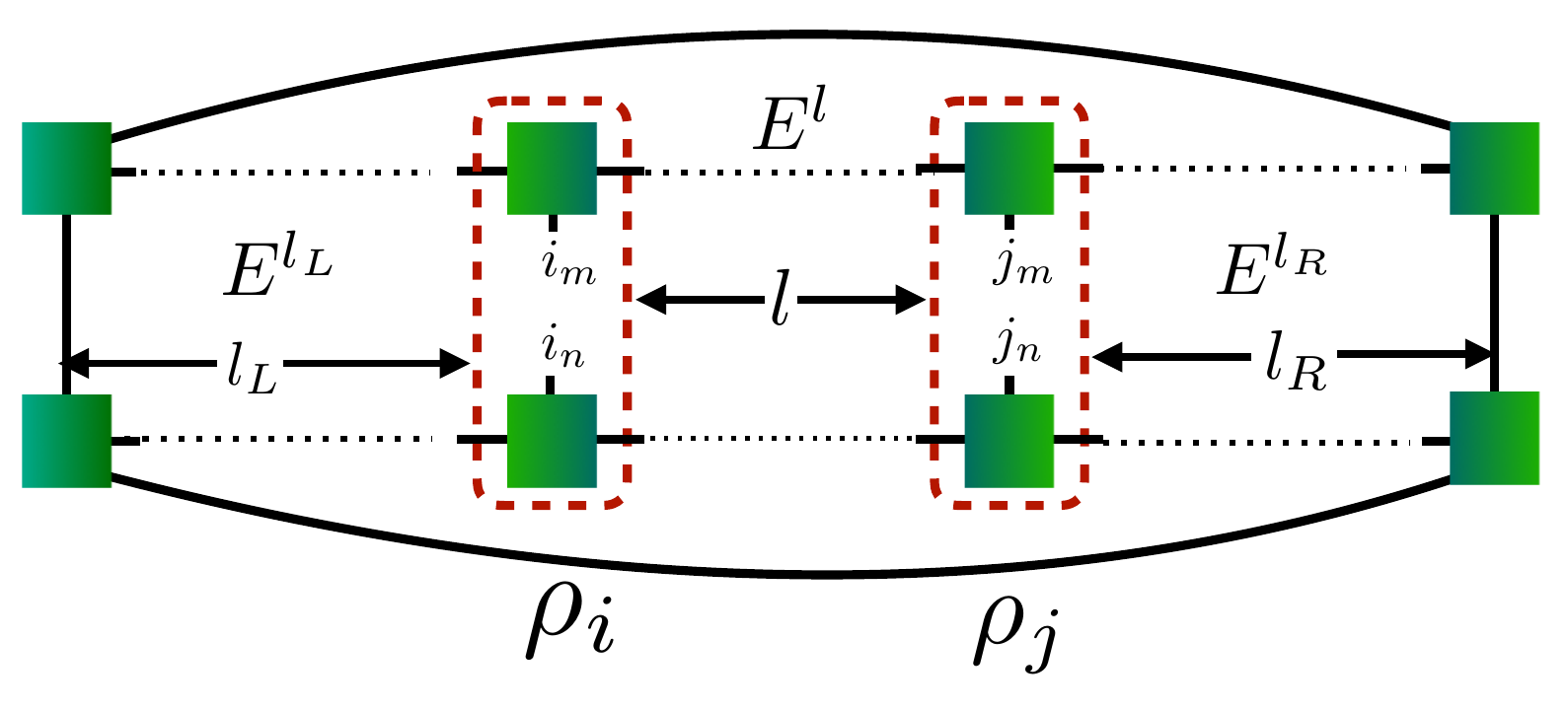}
\caption{Schematic illustration of the few-site reduced density matrix
  obtained from a periodic MPS by contraction of tensors. $\rho_i$ and
  $\rho_j$ denote single-site density matrices which are situated at a
  distance of $l$ sites from each other and $E$ is the transfer
  matrix.}
\label{MPS_schematic}
\end{figure} 

\beq
\rho_{ij}=\Tr_{\overline{\{i,j\}}}\ket|\Psi>\bra<\Psi|.
\eeq
Let us express the above computation in the language of the transfer
matrix $E=\sum_{i_k} A_{i_k}\otimes A^*_{i_k}$, where $A^*$ is the
complex conjugate of $A$, and let us make use of its representations
in left and right eigenvectors $E=\sum_{s=0}^{D^2-1}\gamma_s
\ket|R_s>\bra<L_s|$. The  two-site reduced state can be written as

\beq
\rho_{ij}=\rho_i \otimes \rho_j+\gamma_1^l \bar\rho_{ij},
\label{rho_decomposition}
\eeq
where $\gamma_1$ is the second highest eigenvalue of the transfer
matrix,

\begin{eqnarray}
  \rho_i&=&\sum_{i_{m}j_{m}} \bra<L_0|A_{i_m}\otimes
  A^*_{i_n} \ket|R_0> \ket|i_m>\bra<i_n|, \nonumber\\
  \rho_j&=&\sum_{i_n j_n} \bra<L_0| A_{j_m} \otimes
  A^*_{j_m} \ket|R_0> \ket|j_m>\bra<j_n|,
\label{red_single}
\end{eqnarray}
and we denote (for $l_L+l_R\gg1$, see Fig. \ref{MPS_schematic})

\begin{equation}
  \bar{\rho}_{ij}=\sum_{i_{m}i_n j_{m}j_n}
  \bra<L_0|A_{i_m}\otimes A^*_{i_n}\ket|R_1>
  \bra<L_1|A_{j_m}\otimes A^*_{j_n}\ket|R_0>. \nonumber\\
\end{equation} 
 
(See Appendix \ref{AppendixB} for the derivation.) Note here that the
two matrices $\rho_i$ and $\rho_j$ in Eq. \eqref{red_single} are the
same, but defined in the local Hilbert spaces of the sites $i$ and
$j$.  Moreover, we have assumed the canonical form of the $A_{i_k}$
matrices (right or left canonical), which yields a unique highest
value of $\gamma_s$, $|\gamma_0|=1$ and all other $\gamma_s$'s are
arranged according to the decreasing value of their modulus.

To compute the entropy function $S(\rho_{ij})$, we have to take the
logarithm of $\rho_{ij}$. Let us denote $\rho^0_{ij}=\rho_i \otimes
\rho_j$. Now let us consider the most general case, when $\rho_{ij}^0$
and $\bar\rho_{ij}$ do not commute. In that case, the expansion of the
logarithm goes as follows.

\begin{eqnarray}
  \log(\rho_{ij})&=&\log(\rho^0_{ij}+\gamma_1^l\bar{\rho}_{ij}) \nonumber\\
  &=&\displaystyle{\log(\rho^0_{ij}) + \gamma_1^l
  \int_0^\infty \frac{1}{\rho^0_{ij}+z} \bar\rho_{ij}
  \frac{1}{\rho^0_{ij}+z} dz
  -\gamma_1^{2l} \int_0^\infty \frac{1}{\rho^0_{ij}+z} \bar\rho_{ij}
  \frac{1}{\rho^0_{ij}+z} \bar\rho_{ij} \frac{1}{\rho^0_{ij}+z} dz
  + O(\gamma_1^{2l +1})}.
\end{eqnarray}
The above expansion presents similarities to Dyson's equation for
Green functions. Now using the above expansion we derive the
approximate analytical expression of $J_{ij}^\mut$, given by (see
Appendix \ref{AppendixC} for the complete derivation)

\beq
J_{ij}^\mut=\frac{1}{2}(S(\rho^0_{ij})-S(\rho_{ij}))=\frac{\gamma_1^l}{2}
\Tr \(\bbar \rho_{ij} \log(\rho^0_{ij} )\)+
\frac{\gamma_1^{2l}}{2}
\(\sum_m \frac{1}{2 E_m} |\bbar \rho_{ij}(mm)|^2+
\sum_{m\neq n} \[\frac{E_n}{\(E_m-E_n\)^2} \log\frac{E_n}{E_m}\]
|\bbar\rho_{ij}(mn)|^2\),
\eeq
where $\bbar\rho_{ij}$ is $\bar{\rho}_{ij}$, expressed in the
eigenbasis of $\rho^0_{ij}$, and $E_m$ are the eigenvalues of
$\rho^0_{ij}$. Now using that $\Tr\bar{\rho}_{ij}=0$, one can show
that the coefficient of $\gamma_1^l$, $\Tr\(\bbar\rho_{ij}
\log(\rho^0_{ij})\)=0$. Hence, finally we have

\beq
  J_{ij}^\mut=
  \frac{\gamma_1^{2l}}{2}
  \(\sum_m \frac{1}{2 E_m} \|\bbar\rho_{ij}(mm)\|^2+
  \sum_{m\neq n} \[\frac{E_n}{(E_m-E_n)^2} \log\frac{E_n}{E_m}\]
  |\bbar\rho_{ij}(mn)|^2\).
\label{J_mutual_analytic}
\eeq

Therefore, the above equation implies that for the cases where the
system admits a MPS form, $J_{ij}^\mut$ decays in general as a power
of the second highest eigenvalue of the transfer matrix. The above
relation holds even when the blocks $i$ and $j$ consists of more than
one site. In that case one needs to compute $\rho^0_{ij}$ and
$\bar{\rho}_{ij}$ accordingly. An example of such case is the AKLT
state which exhibits a perfect area law when the blocks $i$, $j$ are
of sufficiently large and in that case $J_{ij}^\mut$ decays as
$(-1/3)^{2l}$. Now one should note here that for a translationally
invariant system, like the MPS considered above,
$J^\mut_{ij}=J^\mut_{i+r,j+r} $ for any $r$. To examine how close the
values of $J_{ij}^\mut$ obtained in Eq. \eqref{J_mutual_analytic}
remain to their exact value $I_{ij}/2$ we have considered the
GS of transverse field Ising model with periodic boundary condition,
given by

\beq
H_{\text{Ising}}=
-\sum_{i=1}^N \sigma_i^x\sigma_{i+1}^x-h_z\sum_{i=1}^N \sigma_i^z,
\label{eq:ITF}
\eeq
where $\sigma^i_k$ are Pauli spin-1/2 operators at site $k$, 
and we compare $J_{ij}^\mut$ and $\frac{I_{ij}}{2}$ for two
different values of the transverse field $h_z$ in Table
\ref{Table1}. 

\begin{table}[h]
\begin{tabular}{|l|l|l|l|l|l|l|l|}
\hline
$h_z$ & $l$ & $I_{1, 2+l}/2$ & $J_{1, 2+l}^\mut $ &$h_z$ & $l$
& $I_{1, 2+l}/2$& $J_{1, 2+l}^\mut$\\
\hline
    & 0 & $0.09776$  & $0.09717$ &     & 0 & $0.05859$ & $0.08080$ \\
    & 1 & $ 0.02693$ & $0.02920$ &     & 1 & $0.00876$ & $0.01161$ \\
1.4 & 2 & $0.00889$  & $0.00881$ & 2.0 & 2 & $0.00150$ & $0.00167$ \\
    & 3 & $0.00310$  & $0.00265$ &     & 3 & $0.00024$ & $0.00024$ \\
    & 4 & $0.00134$  & $0.00079$ &     & 4 & $0.00004$ & $0.00003$ \\ \hline
\end{tabular}
\caption{Comparing the link strengths $J_{ij}^\mut$ obtained using the
  approximate analytical expression derived in
  Eq. \eqref{J_mutual_analytic} to the exact mutual information
  $I_{ij}$ between sites $i$ and $j$, with $l=|j-i-1|$. Here $N=12$.}
\label{Table1}
\end{table}

\subsubsection{Contiguous blocks approximation}

Next, let us estimate the link strengths from the EE of contiguous
blocks, using expression \eqref{eq:blockTI}, which will be denoted by
$J^\cont_r\equiv J_{i,i+r}$. Within this approximation, we  only need to 
know the EE of contiguous blocks, $S(r-1)$, $S(r)$ and $S(r+1)$. As in
the previous case, we will find that $J^\cont_r$ also decays
exponentially with $r$.

We must obtain an analytical expression for the block entropies
$S(r)$. Let us rewrite the general MPS form given in
Eq. \eqref{MPS_general} as

\beq
  \ket|\Psi>=\sum_{\alpha,\beta}
  \ket|\phi_{\alpha\beta}^r> \ket|\phi_{\beta\alpha}^{N-r}>,
\eeq
where $\ket|\phi^r_{\alpha\beta}>=\sum_{i_1i_2\dots
  i_r}\<\alpha|A_{i_1} \dots A_{i_r}|\beta\> |i_1\dots i_r\>$ and
$|\phi^{N-r}_{\beta \alpha}\>=\sum_{i_{r+1}i_{r+2}\dots i_N}
\<\beta|A_{i_{r+1}} \dots A_{i_N}|\alpha\> |i_{r+1}\dots i_N\>$. Now
in general the sets of basis vectors $\{\ket|\phi_{\alpha\beta}^r>\}$
or $\{\ket|\phi_{\beta\alpha}^{N-r}>\}$ are not orthogonal, thus
forcing us to apply Gram-Schmidt's procedure to orthogonalize them in
order to obtain the Schmidt decomposition,

\beq
\ket|\Psi>=
\sum_k \frac{1}{\sqrt{\lambda^r_k\lambda^{N-r}_k}}
\ket|\psi_k^r>\ket|\psi_k^{N-r}>,
\eeq
where $\ket|\psi_k^r>=\sum_{\alpha\beta}c^k_{\alpha\beta}
|\phi^r_{\alpha\beta}\rangle$ and
$|\psi_{k}^{N-r}\rangle=\sum_{\alpha\beta}
d^k_{\alpha\beta}|\phi_{\alpha \beta}^{N-r}\rangle$ with
$\<\psi_{k'}^r|\psi_{k}^r\>=\delta_{kk'}\lambda_k^{r}$,
$\<\psi_{k'}^{N-r}|\psi_{k}^{N-r}\>=\delta_{kk'}\lambda_k^{N-r}$ and
we define $\Lambda(r,N)_k\equiv \lambda_k^{r} \lambda_k^{N-r}$. Thus
the entropy for the $r:N-r$ bipartition is given by

\beq
 S(r)=-\sum_k \Lambda(r,N)_k\log\Lambda(r,N)_k.
 \label{eqn:entropy}
\eeq
Now the Schmidt values $\lambda_k^r$ $\lambda_k^{N-r}$ can be
obtained in terms of the right and left eigenvectors of the transfer
matrix (see Appendix \ref{AppendixD}),
%
and we can finally  write $\Lambda(r,N)_k$ as

\beq
\Lambda(r,N)_k=\lambda_k^r \lambda_k^{N-r} \equiv
\Lambda_k + \sum_s \gamma_s^r \Lambda_k^s,
\eeq
where we denote

\beq
  \Lambda_k = \(\sum_{\substack{\alpha, \beta}}
  |c_{\alpha\beta}|^2\<\alpha\alpha|R_0\>\<L_0|\beta\beta\>\) 
  \(\sum_{\substack{\alpha, \beta}}
  |d_{\alpha\beta}|^2\<\alpha\alpha|R_0\>\<L_0|\beta\beta\>\),
\eeq
and $\Lambda^s_k$ is the coefficient of $\gamma_s^r$ appearing in the
higher order terms. Plugging these in Eq. \eqref{eqn:entropy}, we get

\beq
S(r) = -\sum_k \Lambda_k(r,N) \log\Lambda_k(r,N) =
-\sum_k \(\Lambda_k + \sum_s\gamma_s^r \Lambda_k^s \)
\log\(\Lambda_k +\sum_s\gamma_s^r \Lambda_k^s\).
\eeq
Now expanding the logarithm as $\log
(\Lambda_k+\sum_s\gamma_s^r\Lambda_k^s)\approx
\log(\Lambda_k)+\sum_s\gamma_s^r
\frac{\Lambda_k^s}{\Lambda_k}$, we reach
\begin{eqnarray}
   S(r)  &=& S_0 - \sum_s \gamma_s^r \omega_s
  -\sum_{ss'} \gamma_s^r \gamma_{s'}^r \Gamma_{ss'},
\label{eqn:block_entropies}
\end{eqnarray}
where we have denoted $S_0=-\Lambda_k \log \Lambda_k$, $\omega_s\equiv \sum_k \Lambda_k^s (1 +
\log\Lambda_k)$ and $\Gamma_{ss'}=\sum_k \frac{\Lambda_k^s
  \Lambda_k^{s'}}{\Lambda_k}$. Hence, we can see that all the block
entropies $S(r)$ converge to the thermodynamic value $S_0$ as a power
of the $\gamma_s$.

\medskip

We are now ready to find the analytical expression of the $J_r^\cont$.
Plugging the expression of $S(r)$ in Eq. \eqref{eq:blockTI}, we
finally get

\begin{eqnarray}
  J_r^\cont &=& S(r)-\frac{S(r-1)}{2}-\frac{S(r+1)}{2} \nonumber\\
  &=& -\sum_s \gamma_s^r \omega_s-\sum_{ss'} \gamma_s^r \gamma_{s'}^r \Gamma_{ss'}
  +\frac{1}{2}\sum_s \gamma_s^{r-1}\omega_s+\frac{1}{2}\sum_{ss'}
  \gamma_s^{r-1} \gamma_{s'}^{r-1} \Gamma_{ss'}
  +\frac{1}{2}\sum_s \gamma_s^{r+1} \omega_s+\frac{1}{2}\sum_{ss'}
  \gamma_s^{r+1} \gamma_{s'}^{r+1} \Gamma_{ss'}, \nonumber\\
  &=&-\sum_s \gamma_s^r
  \(1-\frac{\gamma_s}{2}-\frac{\gamma_s^{-1}}{2}\)\omega_s
  - \sum_{ss'} \gamma_s^r \gamma_{s'}^r \(1-\frac{\gamma_s\gamma_{s'}}{2}-\frac{{\gamma_s}^{-1}
  {\gamma_{s}}^{-1}}{2}\)\Gamma_{ss'}.\nonumber\\
\label{eqn:J_continuous_analytic}
\end{eqnarray}
Hence, the link strengths $J_r^\cont$ decay as a power of $\gamma_s$
and for large $r$, eventually decays to zero, indicating area-law
feature in the quantum many-body state.

\medskip

\def\opt{\text{opt}}

Table \ref{Table2} compares the values of $J_r^\cont$ obtained through
Eq.  \eqref{eqn:J_continuous_analytic} to the optimal link
strengths, denoted as $J_r^\opt$. From the
comparison we note that $J_{r}^\cont$ almost coincides with
$J_{r}^{\opt}$ for all $r$ and $h_z$ values.

\begin{table} [h]
\begin{tabular}{ |l|l|l| l|l|l|l|l|}
\hline
$h_z$ & $r$ & $J_r^\opt $& $J_{r}^\cont$ &$h_z$ & $r$ & $J_r^\opt $& $J_{r}^\cont$\\ \hline
    & 1 & $0.09318$ & $0.09776$ &     & 1 & $0.06066$ & $0.05859$ \\
    & 2 & $0.02318$ & $0.02128$ &     & 2 & $0.01006$ & $0.01146$ \\
1.4 & 3 & $0.00647$ & $0.00646$ & 2.0 & 3 & $0.00200$ & $0.00183$ \\
    & 4 & $0.00216$ & $0.00206$ &     & 4 & $0.00042$ & $0.00029$ \\
    & 5 & $0.00097$ & $0.00068$ &     & 5 & $0.00002$ & $0.00004$ \\ \hline
\end{tabular}
\caption{Comparison of the values of $J_r^\cont$ obtained using
  approximate analytical formula for contiguous blocks,
  Eq. \eqref{eqn:block_entropies} with the optimal values, $J_r^\opt$,
  for the Ising model in a transverse field, Eq. \eqref{eq:ITF},
  always using $N=12$.}
\label{Table2}
\end{table}

We may conclude that the link strengths for matrix product states,
when they are approximated using either Eq. \eqref{J_mutual_analytic}
or \eqref{eqn:J_continuous_analytic} decay exponentially with a
certain correlation length associated to the second maximal eigenvalue of the
transfer matrix. In fact, this property may be considered the hallmark
of the area law.


\section{Physical applications of the link representation}
\label{sec:applications}

In this section we apply the link representation formalism to the
analysis of two different physical systems of interest. First of all,
we will consider the different phases of a long-range spin-1/2
Hamiltonian. Then, we will consider the phase diagram of the
bilinear-biquadratic spin-1 Hamiltonian.

\subsection{Long-range Hamiltonian}
\label{results:long_range}

In order to explore the effect of long-range interactions on the
emergent geometry, we start our numerical investigation computing the
link representation for the GS of an interacting Hamiltonian
presenting long-range interaction, thus generalizing results from the
previous work \cite{SinghaRoy_20}, where we considered only similar
systems in the short-range regime. For that purpose, we consider the
long-range spin-1/2 XYZ Hamiltonian where the spin-spin couplings
follow a power-law decay, which can be expressed as

\beq
H=\sum_{i>j} \frac{1}{|i-j|^\alpha}
\(t_x S_i^xS_{j}^x + t_y S^y_iS^y_j + t_z  S_i^zS^z_j\),
\label{Ham:Heisenberg}
\eeq
where $S^k$ are the Pauli spin-1/2 operators $\{k\in x,y,z\}$,
$\alpha>0$ is the interaction exponent, and we consider periodic
boundaries. The model exhibits a very rich phase diagram. In the
ferromagnetic case, $t_x=t_y=-1$, apart from the gapless XY phase and
the ferromagnetic phase that appears in the short-range limit
($\alpha\gg1$), at small values of $\alpha$ ($\alpha \leq 3$) a
gapless continuous symmetry breaking (CSB) emerges \cite{Gorshkov}.
In addition to this, for $t_x=t_y=t_z$ and $\alpha=2$, the model
reduces to the Haldane-Shastry model, which is exactly solvable
\cite{H_Shastry1,H_Shastry2}. Along with this, in recent times,
bipartite \cite{long_range_bipartite} and multipartite
\cite{long_range_multipartite} entanglement studies of the model have
revealed several other interesting properties.

\begin{figure}[h]
\begin{center}
\includegraphics[width=8.0cm]{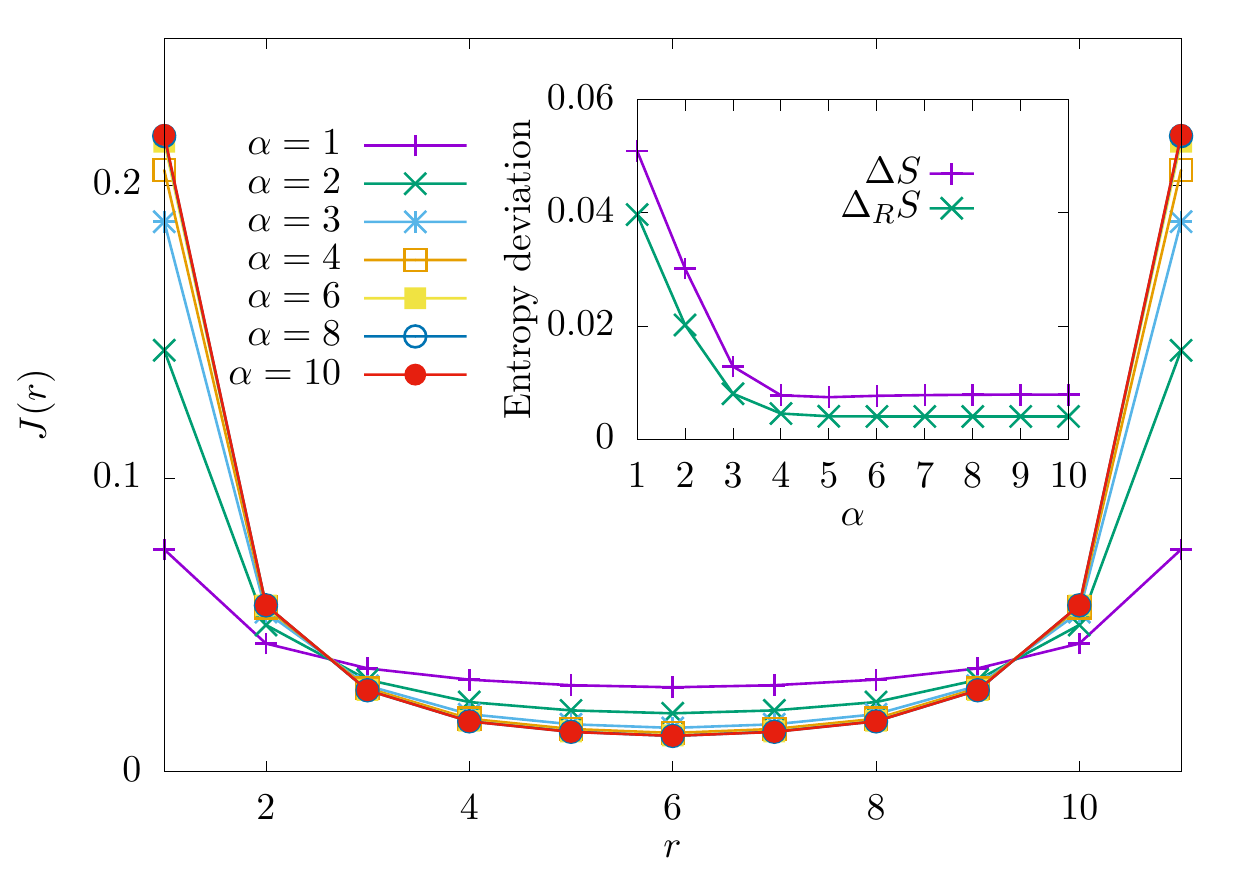}
\includegraphics[width=8.0cm]{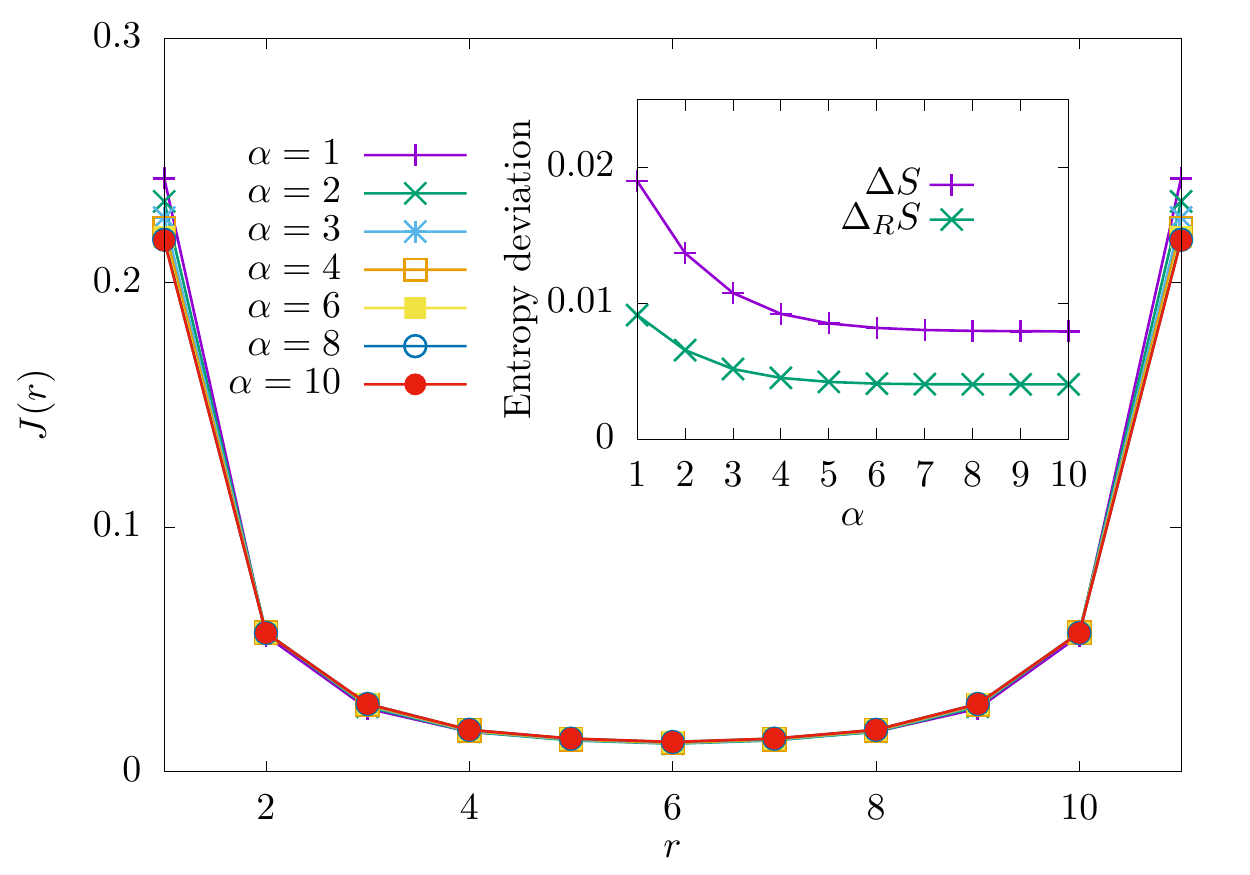}
\caption{Left: decay of $J(r)$ with the site distance $r$ computed for
  the long-range Hamiltonian defined in Eq. \eqref{Ham:Heisenberg} for
  different interaction exponents $\alpha$. Here, we consider the
  interaction in the $XY$ plane to be ferromagnetic ($t_x=t_y=-1$) and
  $t_z=1/2$, for $N=12$. The inset depicts the behavior of the
  absolute error ($\Delta S$) and relative error ($\Delta_RS$) as
  quantified in Eq. \ref{eq:abserror} and Eq. \ref{eq:exprelerror}
  respectively, with the long-range parameter $\alpha$. Right. Same
  plot for an antiferromagnetic interaction in the $XY$ plane,
  $t_x=t_y=1$.}
\label{EAM_long}
\end{center}
\end{figure} 

In the following, we compute the link strengths $J_{ij}$, which reduce
to $J(r)=J_{i,i+r}$ due to the translational invariance, for the
ferromagnetic ($t_x=t_y=-1$) and antiferromagnetic ($t_x=t_y=1$)
interactions in the $XY$ plane, emphasizing the role of long-range
interaction.

\subsubsection{Ferromagnetic interaction in the $XY$ plane: $t_x=t_y=-1$}

We first consider the case when the interaction in the $XY$ plane is
ferromagnetic ($t_x=t_y=-1$) and plot the behavior of $J(r)$ with $r$
for different values of $\alpha$ in Fig. \ref{EAM_long} (left). The
case $\alpha=1$ corresponds to the long-range scenario and the case
$\alpha=10$ to the short-range limit \cite{SinghaRoy_20}. The profiles
of $J(r)$ show a clear dependence on the interaction exponent
$\alpha$. In particular, for $\alpha=1$, $J(r)$ becomes almost flat
for $r\geq2$, which is very different from the short-range
limit. Moreover, we note that this behavior is very much akin to that
obtained for the symmetry-broken antiferromagnetic phase of the model
that arises in the short-range limit ($|t_z|\gg|t_x|=|t_y|$, $\alpha
\gg1$) which we have already explored in \cite{SinghaRoy_20}. Such a
dependence of $J(r)$ on the interaction exponent remains significant
up to $\alpha\leq 3$, i.e. while the system remains in the CSB
phase. Beyond that point, the plots for different values of $\alpha$
correspond to the short range limit. Additionally, we note that both
absolute ($\Delta S$) and relative ($\Delta_R S$) errors of the
entropies are maximum for the perfect long-range case $(\alpha=1)$ and
decays with $\alpha$.

\subsubsection{Antiferromagnetic interaction in $XY$ plane: $t_x=t_y=1$}

We next consider the case when the interaction in the $XY$ plane is
antiferromagnetic, i.e., $t_x=t_y=1$. The behavior of $J(r)$ for
different interaction exponents is shown in Fig. \ref{EAM_long}
(right). From the figure, we can see that unlike the ferromagnetic
case, the $J(r)$ profiles present a very mild dependence on $\alpha$,
and $J(r)$ always decays fast. In the short-range limit, the
ferromagnetic model becomes almost indistinguishable from the
antiferromagnetic one and that is reflected in the identical profiles
of $J(r)$ for $\alpha\approx 10$. In this case, we also note that in
the long-range limit, both absolute error ($\Delta S$) and relative
error ($\Delta_R S$) obtained remain relatively smaller than the
ferromagnetic case.

From these two cases we are led to claim that the GS of Hamiltonian
\eqref{Ham:Heisenberg} presents a 1D entanglement geometry for the
ferromagnetic case with $\alpha\geq 3$ and the anti-ferromagnetic case
for all values of the interaction exponent. Notice that the error of
the link representation is always lower when the entanglement geometry
is well defined. Indeed, the error associated to the optimal link
representation can be considered as a measure of {\em departure} from
a generalized version of the area law.

\subsection{Spin-1 Bilinear-Biquadratic Hamiltonian}
\label{results:spin-1}

We next extend our investigation to a higher spin system, the spin-1
bilinear-biquadratic Heisenberg (BBH) chain with periodic boundaries
 \cite{BBH_1,BBH_2,BBH_3}, which can be expressed as

\beq
H_{BBH}=\sum_i^N \cos(\theta)\; \vec{S}_i\cdot\vec{S}_{i+1} +
\sin(\theta)\; \(\vec{S}_i\cdot\vec{S}_{i+1}\)^2,
\label{Ham:BBH}
\eeq
where $\vec S_i$ are the spin-1 operators. In our case, we mainly
focus on the Haldane phase of the model \cite{Haldane}, which appears
in the range $-\frac{\pi}{4}<\theta<\frac{\pi}{4}$. The GS obtained in
this region is unique and separated by a finite gap from the first
excited state. We obtain the ground state of the model which is a
singlet, using exact diagonalization for different values of $\theta$
and compute the corresponding link matrices from the distribution of
the entanglement in all its possible bipartitions.

\begin{figure}[h]
\begin{center}
\includegraphics[width=8.2cm]{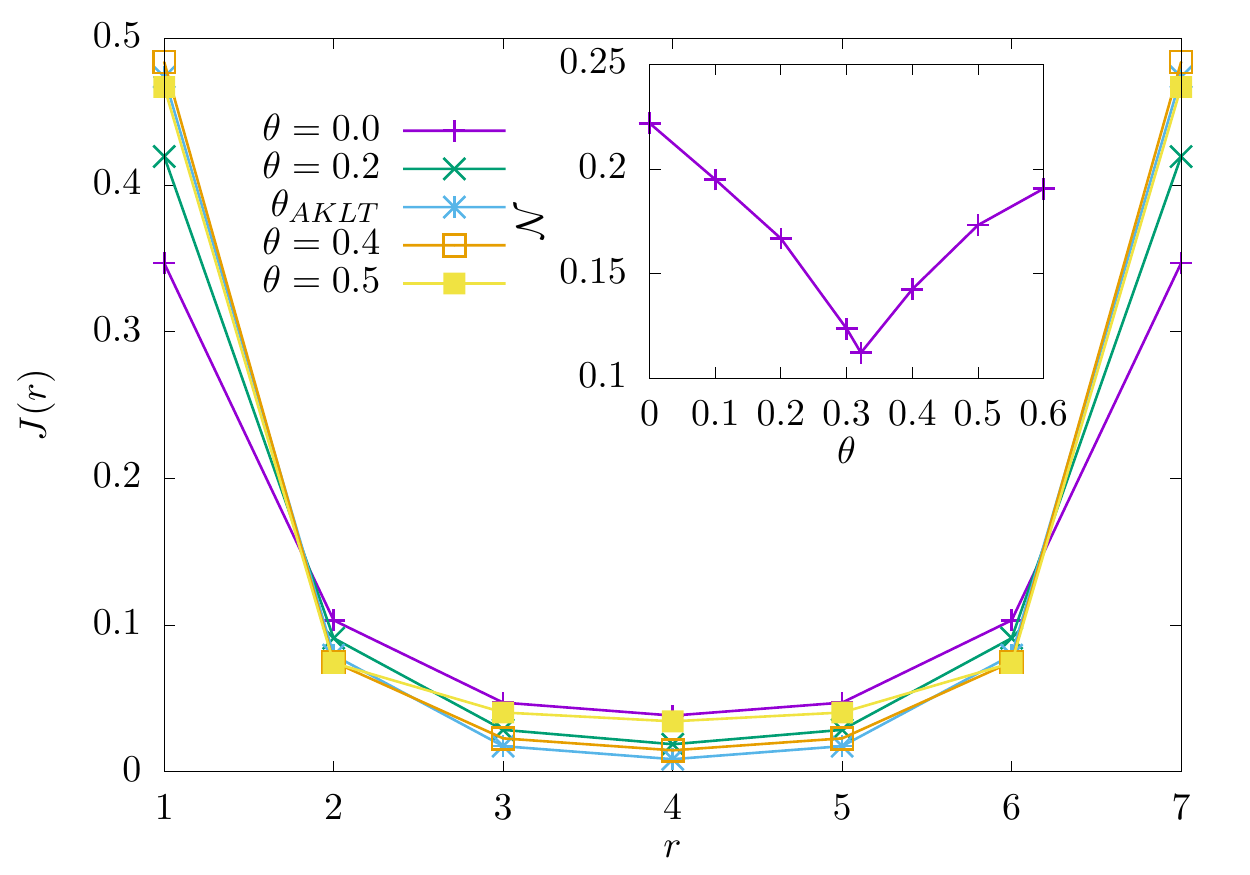}
\includegraphics[width=8.0cm]{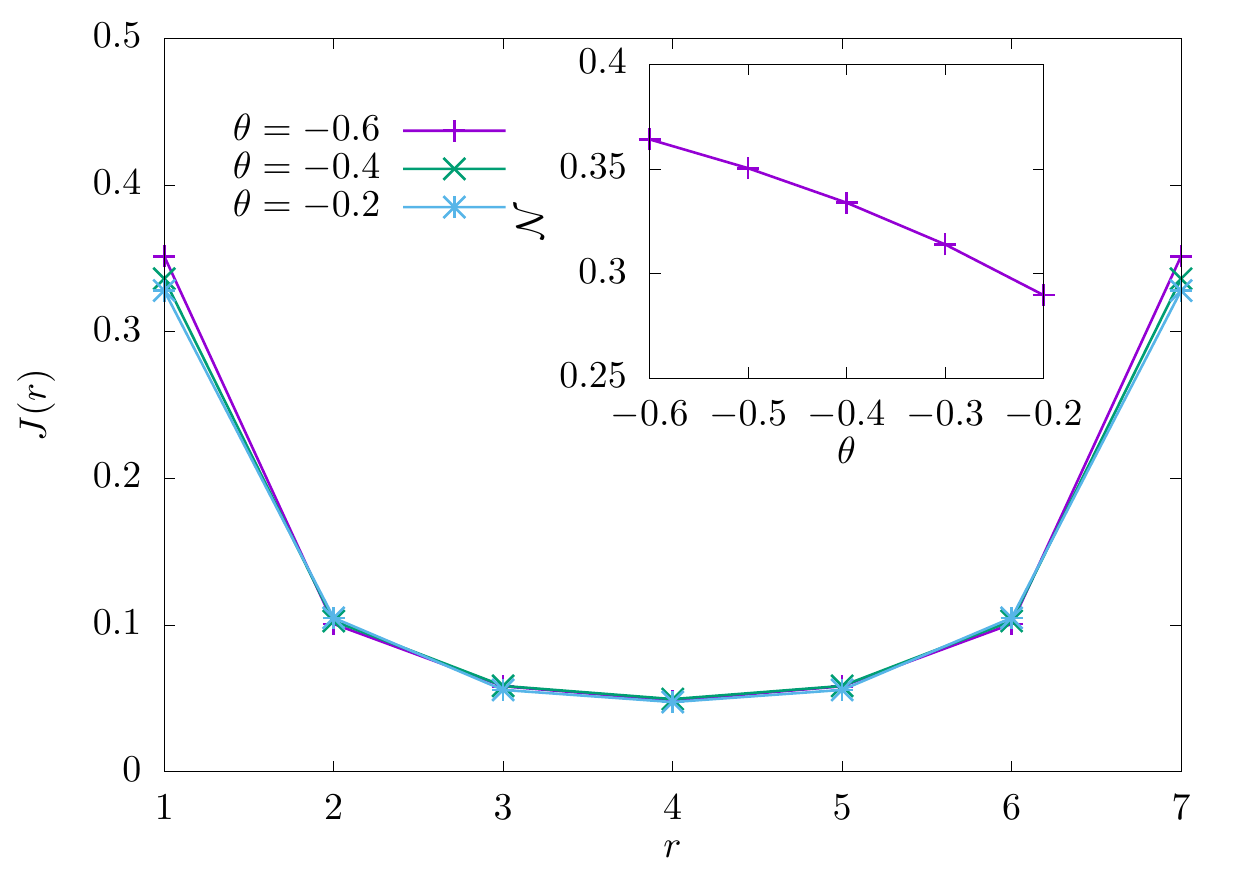}
\caption{Left: Plot of $J(r)$ for the GS of the BBH Hamiltonian
  defined in Eq. \eqref{Ham:BBH} for different values of $\theta$. The
  profile obtained at the AKLT point ($\theta=\arctan(\frac{1}{3})$)
  yields the minimum value of $J(r)$ for $r\geq 2$. In the inset, we
  plot the behavior of the negativity ($\mathcal{N}$) obtained for
  reduced density matrix $\rho_{ij}$ of any two nearest-neighbor sites
  $i,j$ with $\theta$. In all cases, $N=8$. Right: Same plot for
  negative values of $\theta$.}
\label{EAM_BBH}
\end{center}
\end{figure} 

\subsubsection{$\theta \ge0$ region}

We first consider the positive $\theta$ region of the Haldane phase
and obtain the behavior of the link strengths, $J(r)$. The profiles
obtained for different $\theta$ values are shown in Fig. \ref{EAM_BBH}
(left). From the figure, one can observe that there is a clear
dependence of $J(r)$ on $\theta$. For $r=1$, the link strength $J(r)$
increases with $\theta$, while for $r\geq2$, it exhibits the opposite
behavior and for any fixed value of $r$ ($2\leq r \leq N/2$), the
value of $J(r)$ obtained for
$\theta=\theta_{AKLT}=\arctan(\frac{1}{3})$ turns out to be
minimum. The reason is that AKLT is actually the fixed point of a real
space renormalization group, so one expects that all entanglement will
be concentrated in the nearest neighbour sites.

This behavior is consistent with the entanglement properties explored
in this region. For example, it is known that the entanglement
negativity, defined as

\beq
\mathcal{N}(\rho)=\frac{|\rho_{ij}|^{T_j}-1}{2},
\label{eq:negativity}
\eeq
where $T_j$ is the partial transpose of the reduced system $\rho_{ij}$
with respect to subsystem $j$ and $|.|$ is the trace norm, exhibits a
local minimum at the AKLT point \cite{BBH_3}, shown at the inset of
Fig. \ref{EAM_BBH} (left).

\subsubsection{$\theta<0$ region}

The behavior of link representation obtained for the $\theta<0$ region
of the Haldane phase is very different from that obtained for
$\theta\ge 0$. We found that in this region, the link strengths $J(r)$
present a very mild dependence on $\theta$ only for $r=1$ and become
independent of $\theta$ from $r\geq2$. Hence, all the $J(r)$ profiles
seem to collapse, as shown in Fig. \ref{EAM_BBH} (right). Similarly to
the positive $\theta$ region, this behavior is also found to be
consistent with the behavior of entanglement in this region
\cite{BBH_3}. In particular, the negativity, $\mathcal{N}$, obtained
for two neighboring sites shows a slow decay with $\theta$. For
completeness, we also plot the behavior of $\mathcal{N}$ with $\theta$
in the inset of Fig. \ref{EAM_BBH} (right).

\begin{figure}[h]
\begin{center}
\includegraphics[width=9.0cm]{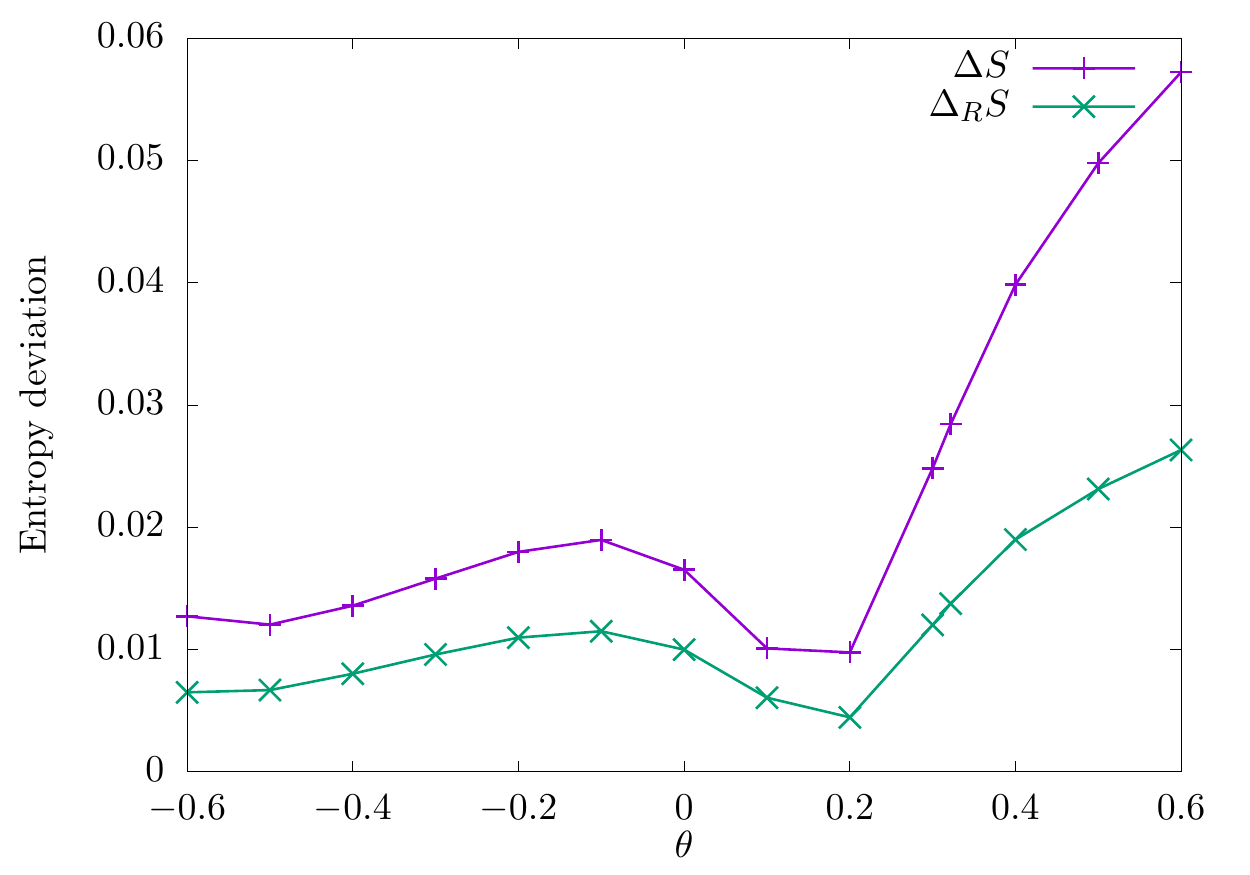}
\caption{Error analysis for BBH Hamiltonian as defined in
  Eq. (\ref{Ham:BBH}). We compare both absolute ($\Delta S$) and
  relative ($\Delta_R S$) error for different values of $\theta$. Here
  N=8.}
\label{fig:error_BBH} 
\end{center}
\end{figure}

Fig. \ref{fig:error_BBH} shows the absolute ($\Delta S$) and relative
($\Delta_R S$) error, as defined in Eq. \eqref{eq:abserror} and Eq.
\eqref{eq:exprelerror} respectively, for the link representation of
the GS of the BBH Hamiltonian as a function of $\theta$. We can
observe that the error remains relatively low, $\Delta S\sim 0.02,
\Delta_R S\sim 0.01$ for $\theta<0.2$, and grows substantially above
that point.

\section{Conclusions and further work}
\label{sec:conclusions}

The entanglement entropies of all possible bipartitions of a pure
state of a quantum system can be expressed in a simple form via the
{\em link representation}, Eq. \eqref{eq:linkrep}, i.e. the entropy of
any block can be estimated summing the {\em link strengths} between
sites of the block and sites of the environment. In \cite{SinghaRoy_20}
a procedure to obtain the optimal link representation of a given
quantum state was presented, along with a proof that the most relevant
properties of entanglement (e.g. strong subadditivity) were naturally
fulfilled.

In this article we provide an analysis of the accuracy of the link
representation for different types of states, focusing on a
conformally invariant free-fermionic state and on random states. The
accuracy of the link representation is very high for the conformally
invariant state, and increases with the system size. For the random
state, on the other hand, we observe that the accuracy is much worse,
showing that the link representation is valuable when the system
presents some effective area law, even if logarithmic corrections are
required.

The high computational cost of obtaining the optimal link
representation suggests the search of effective approximate
methods. We provide several which provide a very good
accuracy. Instead of considering the entanglement entropies for all
bipartitions we may restrict ourselves to a random sample, and we show
that a number of samples corresponding to a few times the number of
parameters is enough for a good accuracy. Moreover, we provide an
approximate link representation based on the knowledge of compact
blocks, and another one specially suited for free-fermionic
states.

We have also considered the link representation of matrix product
states (MPS). We have shown that the link strengths for an MPS decay
exponentially with the distance, with the correlation length
corresponding to the inverse of the second highest eigenvalue of the transfer
matrix. In order to prove that result, we have computed the mutual
information of pairs of sites and the entropy of contiguous blocks in
generic MPS, and applied them to the evaluation of the link strengths
for the off-critial Ising model in a transverse field.

We have obtained the link representation of the GS of two relevant
physical models, in order to check how the accuracy and the link
strengths signal the presence of different quantum phases. For the
Heisenberg model with long-range interactions, we have found that the
accuracy of the link representation becomes higher in the short-range
phase, which shows that this accuracy can be used to signal departure
from the area law. The behavior of the link strengths is also
different in the long-range phase, becoming homogeneous, as they do
for random states. Also, we have considered the spin-1
bilinear-biquadratic Hamiltonian in the vicinity of the AKLT point,
showing that the representation of the entanglement entropies of the
state is good, as they should because the states can be represented as
MPS of low bond dimension.


\begin{acknowledgments}
We would like to thank P. Calabrese, J.I. Cirac, J.I. Latorre, E.
López, L. Tagliacozzo, E. Tonni, G. Vidal, H.Q. Zhou, Q.Q. Shi and
S.Y. Cho for conversations.  We acknowledge financial support from the
grants PGC2018-095862-B-C21, PGC2018-094763-B-I00,
PID2019-105182GB-I00, QUITEMAD+ S2013/ICE-2801, SEV-2016-0597 of the
{\em Centro de Excelencia Severo Ochoa} Programme and the CSIC
Research Platform on Quantum Technologies PTI-001.
\end{acknowledgments}


\newpage
\appendix

\section{Graphical proof of $(\Delta_2 \Delta_1 S)_{i,j}=2J_{ij}$}
In this section we present a graphical proof of Eq. (\ref{eq:D2S}) from the main text. The black plus signs correspond to the entanglement links which we must add to obtain $S_{i,j}$, while the red plus signs and the green and blue minus signs correspond to the other terms in the equation. Indeed, we can easily check that all plus and minus signs cancel out except for the element $(i,j)$, which contains two plus signs, as claimed in  Eq. (\ref{eq:D2S}).
\begin{figure}[h]
\label{AppendixA}
  \includegraphics[width=9cm]{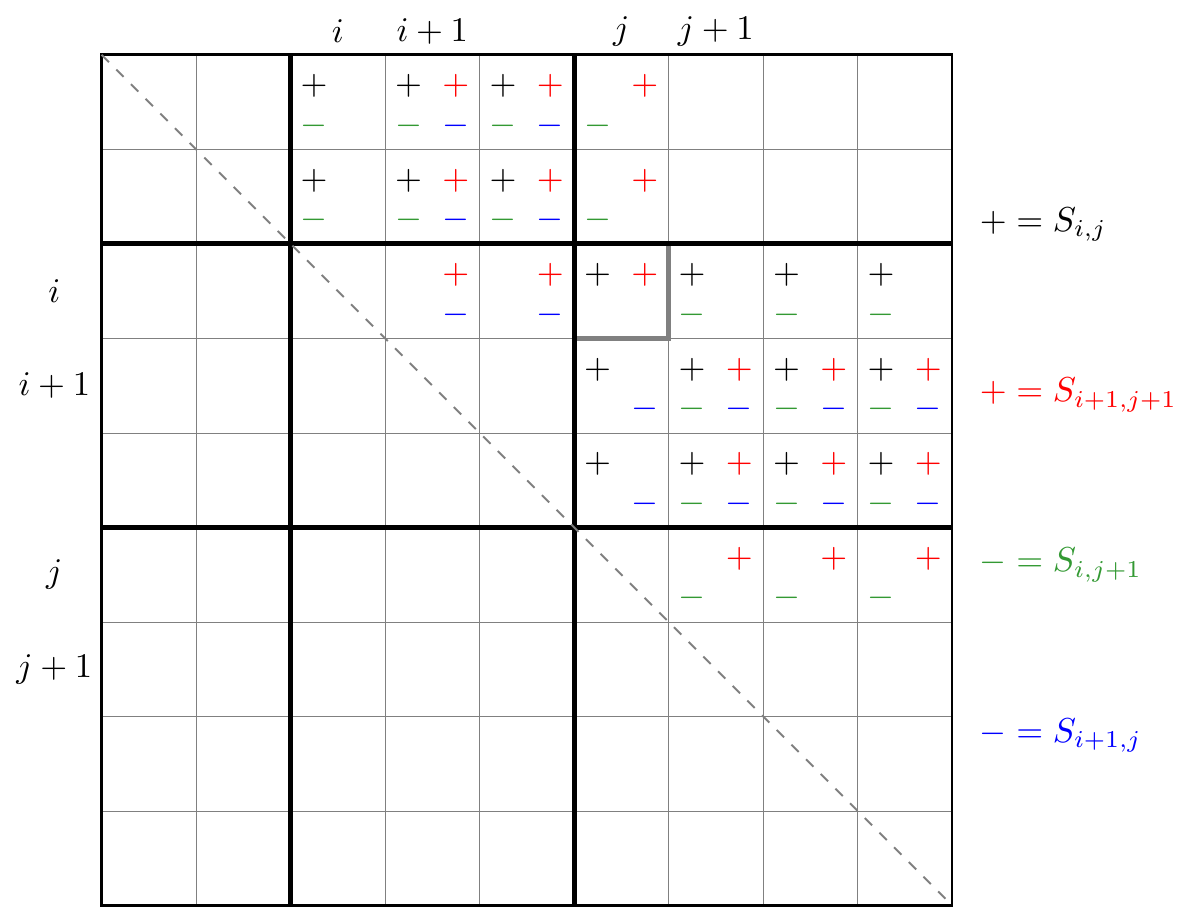}
  \caption{Graphical proof of Eq. \eqref{eq:D2S}. We show the full
    $J$-matrix, highlighting the the entanglement links contributing
    to $S_{i,j}$ (black, $+$), $S_{i+1,j+1}$ (red, $+$), $S_{i+1,j}$ (blue $-$)  and
    $S_{i,j+1}$ (green, $-$). Notice that all entanglement links cancel
    out except $J_{i,j}$, which gets two positive signs.}
  \label{fig:d2s}
\end{figure}

\section{Few-sites reduced density matrix for MPS}
\label{AppendixB}

We start with a matrix product state representation as given in
Eq. \eqref{MPS_general} in the main text

\beq
|\Psi\>=\sum_{i_1 i_2 \dots i_N} \text{Tr}(A_{i_1} A_{i_2}\dots A_{i_N}) |i_1 i_2\dots i_N\>.
\eeq

Now consider the reduced density matrix $\rho_{ij}$, which can be
obtained from $|\psi\>$ by tracing out all but the sites $i$ and
$j$, as follows

\beq
\rho_{ij}=\Tr_{\overline{\{i,j\}}}\ket|\Psi>\bra<\Psi|.
\eeq
Now from the schematic in Fig. \ref{MPS_schematic} of the main text,
it is easy to realize

\beq
 \rho_{ij}=\Tr(E^{l_L} \bar{E}_i E^l \bar{E}_j E^{l_R}),
\eeq
where we consider the system is translational invariant and
$\bar{E}_k=\sum_{k_m k_n} (A_{k_m} \otimes A^*_{k_n})|k_m \>
\< k_n|$. Expanding $E= |R_0\> \< L_0|+\sum_s \gamma_s
|R_s\> \< L_s|$, we get

\begin{eqnarray}
{\rho}_{ij}&=&\Tr\[\(|R_0\>\< L_0|+\sum_q\gamma_q^{l_L} |R_q\>\<L_q|\)
\bar{E}_i \(|R_0\>\<L_0|+\gamma_1^{l}|R_1\>\< L_1|\)
\bar{E}_j \(|R_0\>\<L_0|+\sum_p\gamma_p^{l_R}|R_p\>\< L_p|\)\],\nonumber\\
&=&\Tr\(|R_0\>\<L_0|\bar{E}_i|R_0\>\< L_0|\bar{E}_j|R_0\>\<L_0|\)+
\Tr\(\gamma_1^l|R_0\>\<L_0|\bar{E}_i|R_1\>\<L_1|\bar{E}_j|R_0\>\<L_0|\)
\nonumber\\
&+&
\Tr\(\sum_{pq}\gamma_q^{l_L}\gamma_p^{l_R}|R_q\>\<L_q|\bar{E}_i|R_0\>
\<L_0|\bar{E}_j|R_p\>\< L_p|\) +
\Tr\(\sum_{pq} \gamma_q^{l_L}\gamma_p^{l_R} \gamma_s^l|R_q\>
\<L_q|\bar{E}_i|R_1\>\<L_1|\bar{E}_j|R_p\>\< L_p|\),\nonumber\\
&=&\<L_0|\bar{E}_i|R_0\> \<L_0|\tilde{E}_j|R_0\> +
\gamma_1^l \<L_0|\bar{E}_i|R_1\>\<L_1|\bar{E}_j|R_0\>+
\sum_{pq}\delta_{pq}\gamma_q^{l_L}\gamma_p^{l_R}
\<L_q|\bar{E}_i|R_0\>\<L_0|\bar{E}_j|R_p\>
\nonumber\\
&+& \sum_{pq} \delta_{pq}\gamma_q^{l_L}\gamma_p^{l_R} \gamma_s^l \< L_q|\bar{E}_i|R_1\> \< L_1|\bar{E}_j|R_p\>, \nonumber\\
&=&\<L_0|\bar{E}_i|R_0\> \< L_0|\bar{E}_j|R_0\>+ \gamma_1^l \<
L_0|\bar{E}_i|R_1\>\<
L_1|\bar{E}_j|R_0\>+\sum_{q}\ \gamma_q^{l_L+l_R} \<
L_q|\bar{E}_i|R_0\> \< L_0|\bar{E}_j|R_q\> \nonumber\\
&+& \sum_{qs} \gamma_q^{l_L+l_R} \gamma_s^l \< L_q|\bar{E}_i|R_s\> \< L_s|\bar{E}_j|R_q\>. \nonumber\\
\end{eqnarray}
Using $l_L+l_R\gg1$, we get 

\begin{eqnarray}
\rho_{ij}&=&\sum_{i_mi_nj_mj_n}\< L_0|A_{i_m} \otimes A^*_{i_n}|R_0\>
\< L_0|A_{j_m} \otimes A^*_{j_n}|R_0\> |i_m j_m\>\<i_nj_n| \nonumber\\
&+&\gamma_1^l \sum_{i_m i_n j_m j_n } \< L_0|A_{i_m} \otimes A^*_{i_n}|R_1\>\< L_1|A_{j_m} \otimes A^*_{j_n}|R_0\>  |i_m j_m\>\< i_n j_n|,\nonumber\\
&=&\rho_i \otimes \rho_j+ \gamma_1^l \tilde{\rho}_{ij},\nonumber\\
\end{eqnarray}
with $\rho_i \otimes \rho_j=\sum_{i_mi_nj_mj_n}\< L_0|A_{i_m} \otimes
A^*_{i_n}|R_0\> \< L_0|A_{j_m} \otimes A^*_{j_n}|R_0\> |i_m j_m\>\<
i_n j_n|$, and $\tilde{\rho}_{ij}=\sum_{i_m i_n j_m j_n} \<
L_0|A_{i_m} \otimes A^*_{i_n}|R_1\>\< L_1|A_{j_m} \otimes
A^*_{j_n}|R_0\>  |i_mj_m\>\<i_nj_n|$.


\section{Mutual information between sites for MPS}
\label{AppendixC}

We start with the expansion of the logarithm

\beq
\log(\rho_{ij})=\log(\rho^0_{ij}+\gamma_1^l\bar{\rho}_{ij})=
\log(\rho^0_{ij}) +
\gamma_1^l \int_0^\infty \frac{1}{\rho^0_{ij}+z} \bar{\rho}_{ij}
\frac{1}{\rho^0_{ij}+z} dz
-\gamma_1^{2l} \int_0^\infty \frac{1}{\rho^0_{ij}+z}
\bar{\rho}_{ij}
\frac{1}{\rho^0_{ij}+z}
\bar{\rho}_{ij}
\frac{1}{\rho^0_{ij}+z} dz
+ O(\gamma_1^{2l +1}).
\eeq
Hence, 

\begin{eqnarray}
  -\rho_{ij}\log(\rho_{ij})&=&-(\rho^0_{ij}+\gamma_1^l\bar{\rho}_{ij})
  \log(\rho^0_{ij}+\gamma_1^l \bar{\rho}_{ij}),\nonumber\\ 
&=&-(\rho^0_{ij}+\gamma_1^l\bar{\rho}_{ij})\log(\rho^0_{ij})
  - \gamma_1^l  (\rho^0_{ij}+\gamma_1^l \bar{\rho}_{ij})
  \int_0^\infty \frac{1}{\rho^0_{ij}+z} \bar{\rho}_{ij}
  \frac{1}{\rho^0_{ij}+z} dz, \nonumber\\
&+&\gamma_1^{2l} (\rho^0_{ij}+\gamma_1^l \bar{\rho}_{ij})
  \int_0^\infty \frac{1}{\rho^0_{ij}+z} \bar{\rho}_{ij}
  \frac{1}{\rho^0_{ij}+z}\bar{\rho}_{ij} \frac{1}{\rho^0_{ij}+z} dz+ O(\gamma_1^{2l+1}),\nonumber\\
&=&-\rho^0_{ij} \log(\rho^0_{ij})-\gamma_1^l\[\bar{\rho}_{ij}
  \log(\rho^0_{ij})+\rho^0_{ij} \int_0^\infty
  \frac{1}{(\rho^0_{ij}+z)}\bar{\rho}_{ij} \frac{1}{(\rho^0_{ij}+z)}
  dz \] \nonumber\\
&-& \gamma_1^{2l}\[ \bar{\rho}_{ij}
  \int_0^\infty \frac{1}{(\rho^0_{ij}+z)} 
  \bar{\rho}_{ij}  \frac{1}{(\rho_0+z)}dz 
  - \rho^0_{ij}\int_0^\infty \frac{1}{(\rho^0_{ij}+z)}
  \bar{\rho}_{ij} \frac{1}{(\rho^0_{ij}+z)} \bar{\rho}_{ij}
  \frac{1}{(\rho^0_{ij}+z)} dz\] + O(\gamma_1^{2l+1}).\nonumber\\
\end{eqnarray}
Let us now consider the terms in the integral.

\begin{eqnarray}
  S(\rho_{ij})&=&S(\rho^0_{ij})-\gamma_1^l
  \[\Tr(\bar{\rho}_{ij}  \log\rho^0_{ij})+
  \Tr\(\rho^0_{ij} \int_0^\infty \frac{1}{\rho^0_{ij}+z}
  \bar{\rho}_{ij}\frac{1}{\rho^0_{ij}+z} dz\) \]
-  \gamma_1^{2l}\[ \Tr  \(\bar{\rho}_{ij}   \int_0^\infty
\frac{1}{\rho^0_{ij}+z} \bar{\rho} \frac{1}{\rho^0_{ij}+z}dz\) \right.\nonumber\\
&-&\left.\Tr \( \rho^0_{ij} \int_0^\infty \frac{1}{\rho^0_{ij}+z} \bar{\rho}_{ij}   \frac{1}{\rho^0_{ij}+z} \bar{\rho}_{ij}  \frac{1}{\rho^0_{ij}+z} dz\)\].
\end{eqnarray}

\medskip

{\bf 1. First term:} $\Tr\(\rho^0_{ij} \int_0^\infty
\frac{1}{\rho^0_{ij}+Z} \bar{\rho}_{ij} \frac{1}{\rho^0_{ij}+z} dz\):$ 
Consider $\rho^0_{ij}$ has eigenvalues $E_m$. Hence, the eigenvalue of
$(\rho^0_{ij}+z)^{-1}$ is given by $(E_m+z)^{-1}$.  Now, if we express
everything in the basis of $\rho^0_{ij}$, we would get

\beq
\rho^0_{ij} \frac{1}{\rho^0_{ij}+z} \bar{\rho}_{ij} \frac{1}{\rho^0_{ij}+z} =\sum_m E_m  (E_m+z)^{-2} \bbar \rho_{ij}(mm),
\label{eq_initial}
\eeq
where $\bbar\rho_{ij}$ is ${\bar{\rho}}_{ij}$, expressed in the
eigebasis of $\rho^0_{ij}$ and $E_{m}$ are eigenvalues of $\rho_0$.
If we integrate \ref{eq_initial}, we would get 

\begin{eqnarray}
  \int_0^\infty  \sum_m E_m  (E_m+z)^{-2} \bbar \rho_{ij}(mm) dz
  &=& \sum_m E_m \bbar \rho_{ij}(mm)\int_0^\infty (E_m+z)^{-2}dz, \nonumber\\
  &=&-\sum_m E_m\[\frac{1}{E_m+z}\]_0^\infty\bbar \rho_{ij}(mm) =\sum_m \bbar
\rho_{ij}(mm)=0 . 
\end{eqnarray}

Hence, we finally get $\Tr\(\rho^0_{ij} \int_0^\infty \frac{1}{\rho^0_{ij}+Z} \bar{\rho}_{ij} \frac{1}{\rho^0_{ij}+z} dz\)=0.$

\medskip

{\bf 2. Second term:} $\Tr \(\bar{\rho}_{ij} \int_0^\infty
\frac{1}{\rho^0_{ij}+Z} \bar{\rho}_{ij} \frac{1}{\rho^0_{ij}+z}dz\)$:
We can again proceed as before and show that

\begin{eqnarray}
\Tr\(\bar{\rho}_{ij}   \frac{1}{\rho^0_{ij}+z} \bar{\rho}_{ij} \frac{1}{\rho^0_{ij}+z}\)=\sum_{mn}  \bbar \rho_{ij}(mn) \bbar \rho_{ij}(nm)   (E_m+z)^{-1}   (E_n+z)^{-1}.
\end{eqnarray}

We will integrate the above term later.

{\bf 3. Third term:} $-\Tr \( \rho^0_{ij} \int_0^\infty \frac{1}{\rho^0_{ij}+z} \bar{\rho}_{ij} \frac{1}{\rho^0_{ij}+z} \bar{\rho}_{ij} \frac{1}{\rho^0_{ij}+z} \)$.

This term is very similar to previous one, except, we have here additional factor $\rho^0_{ij} \frac{1}{\rho^0_{ij}+z}$, which is   diagonal in the basis of $\rho^0_{ij}$, and when multiplied with rest of the term $ \bar{\rho}_{ij} \frac{1}{\rho^0_{ij}+z} \bar{\rho}_{ij} \frac{1}{\rho^0_{ij}+z} $ just contributes $E_m (E_m+z)^{-1}$ in the diagonal.

Hence, if we sum second and third term. we would get

\begin{eqnarray}
&&\Tr  \(\bar{\rho}_{ij}   \frac{1}{\rho^0_{ij}+z} \bar{\rho}_{ij} \frac{1}{\rho^0_{ij}+z}\)-\Tr \( \rho^0_{ij} \frac{1}{\rho^0_{ij}+z} \bar{\rho}_{ij} \frac{1}{\rho^0_{ij}+z} \bar{\rho}_{ij} \frac{1}{\rho^0_{ij}+z} \), \nonumber\\
&=&\sum_{mn}  \bbar \rho_{ij}(mn) \bbar \rho_{ij}(nm)     (E_m+z)^{-1}  (E_n+z)^{-1}  ( 1-E_m (E_m+z)^{-1} )\nonumber,\\
&=&\sum_{mn}  \bbar \rho_{ij}(mn) \bbar \rho_{ij}(nm)    z  (E_m+z)^{-2}  (E_n+z)^{-1}. 
\end{eqnarray}

We will  now integrate the above term,
\begin{eqnarray}
\Tr  \(\bar{\rho}_{ij} \int_0^\infty \frac{1}{\rho^0_{ij} +z}
\bar{\rho} \frac{1}{\rho^0_{ij} +z}dz\)-\Tr \( \rho^0_{ij}
\int_0^\infty \frac{1}{\rho^0_{ij} +z} \bar{\rho}_{ij}
\frac{1}{\rho^0_{ij} +z} \bar{\rho}_{ij}  \frac{1}{\rho^0_{ij} +z}
dz\)\nonumber\\
=\sum_{mn}  \bbar \rho_{ij}(mn) \bbar \rho_{ij}(nm)    \int_0^\infty  z  (E_m+z)^{-2}  (E_n+z)^{-1} dz.
\label{eq0}
\end{eqnarray}

Now for $m\neq n$, let us write $ z  (E_m+z)^{-2}  (E_n+z)^{-1} =\frac{E_m}{E_m-E_n}  \frac{1}{(E_m+z)^2}+\frac{E_n}{(E_m-E_n)^2} \frac{1}{E_m+z}-\frac{E_n}{(E_m-E_n)^2} \frac{1}{E_n+z}$. Hence, integrating, we get

\begin{eqnarray}
\int_0^\infty z  (E_m+z)^{-2}  (E_n+z)^{-1} &=&\int_0^\infty\frac{E_m}{E_m-E_n}  \frac{dz}{(E_m+z)^2}+\frac{E_n}{(E_m-E_n)^2} \frac{dz}{E_m+z}-\frac{E_n}{(E_m-E_n)^2} \frac{dz}{E_n+z},\nonumber\\
&=&-\frac{E_m}{E_m-E_n} \[\frac{1}{E_m+z}\]_0^\infty+
\frac{E_n}{(E_m-E_n)^2} \[\log\(\frac{E_m+z}{E_n+z}\)\]_0^\infty,\nonumber\\
&=&\frac{1}{E_m-E_n} +\frac{E_n}{(E_m-E_n)^2} \log\(\frac{E_n}{E_m}\).
\label{eq1}
\end{eqnarray}

Similarly, for $m=n$, we have 
\beq
 \int_0^\infty z  (E_m+z)^{-3}
 dz=\int_0^\infty \frac{dz}{(E_m+z)^2} -\int_0^\infty\frac{E_m}{(E_m+z)^3}dz=\[-\frac{1}{(E_m+z)}+\frac{E_m}{2(E_m+z)^2}\]_0^{\infty} 
 =\frac{1}{2E_m}.
 \label{eq2}
\eeq

Combining Eqs. (\ref{eq1}) and (\ref{eq2}), we finally get
\begin{eqnarray}
&&\Tr  \(\bar{\rho}_{ij}    \int_0^\infty \frac{1}{\rho^0_{ij} +z} \bar{\rho}_{ij}  \frac{1}{\rho^0_{ij} +z}dz\)-\Tr \( \rho^0_{ij}  \int_0^\infty \frac{1}{\rho^0_{ij} +z} \bar{\rho}_{ij}  \frac{1}{\rho^0_{ij} +z} \bar{\rho}_{ij}  \frac{1}{\rho^0_{ij} +z} dz\)\nonumber\\
&=&\sum_m \frac{1}{2 E_m}   |\bbar \rho_{ij}(mm)|^2+\sum_{m\neq n} \[\frac{1}{E_m-E_n} +\frac{E_n}{(E_m-E_n)^2} \log\(\frac{E_n}{E_m}\)\] |\bbar \rho_{ij}(mn)|^2.
\end{eqnarray}
Now  $|\bbar \rho_{ij}^2|$ remains unchanged under the  exchange of  indices $m$ and $n$ but $\frac{1}{E_m-E_n}$ acquires a minus sign. Hence, $\sum_{m,n}\frac{1}{E_m-E_n}  |\bbar \rho_{ij}(mn)|^2=0.$
Therefore, we can finally write

\beq
S(\rho)=S(\rho_0)-\gamma_1^l\Tr \(\bbar \rho_{ij} \log(\rho^0_{ij} )\)-\gamma_1^{2l}\(\sum_m \frac{|\bbar \rho_{ij}(mn)|^2}{2 E_m}  +\sum_{m\neq n} \[\frac{E_n}{(E_m-E_n)^2} \log\(\frac{E_n}{E_m}\)\] |\bbar \rho_{ij}(mn)|^2\).
\eeq

\begin{eqnarray}
J_{ij}^\mut
=\frac{1}{2}(S(\rho^0_{ij})-S(\rho_{ij}))=\frac{\gamma_1^l}{2} \Tr
\(\bbar \rho_{ij} \log(\rho^0_{ij} )\)+\frac{\gamma_1^{2l}}{2}\(\sum_m
\frac{|\bbar \rho_{ij}(mm)|^2}{2 E_m} 
+\sum_{m\neq n} \[\frac{E_n}{(E_m-E_n)^2} \log\(\frac{E_n}{E_m}\)\] |\bbar \rho_{ij}(mn)|^2\).\nonumber\\
\end{eqnarray}


\section{Schmidt values for MPS}
\label{AppendixD}

From the expression of $|\psi_k^r\>=\sum_{\alpha\beta}c_{\alpha\beta}|\phi^r_{\alpha\beta}\>$ and $|\psi_k^{N-r}\>=\sum_{\alpha\beta} d_{\alpha\beta}|\phi^{N-r}_{\alpha\beta}\>$, with 
$|\phi^r_{\alpha\beta}\>=\sum_{i_1i_2\dots i_r}\< \alpha|A_{i_1} \dots
A_{i_r}|\beta\> |i_1\dots i_r\>$ and $|\phi^{N-r}_{\beta
  \alpha}\>=\sum_{i_{r+1}i_{r+2}\dots i_{N}}\< \beta|A_{i_{r+1}} \dots
A_{i_N}|\alpha\> |i_{r+1}\dots i_N\>$ we can obtain the Schmidt value
as follows

\begin{eqnarray}
  \lambda_k^r=\<\psi_{k}^r|\psi_{k}^r\>
  &=& \sum_{\alpha\alpha'\beta\beta'} c_{\alpha\beta}
  c^*_{\alpha'\beta'}
  \<\phi^r_{\alpha'\beta'}|\phi^r_{\alpha \beta}\>, \nonumber\\
  &=& \sum_{\alpha=\alpha',\beta=\beta'} |c_{\alpha\beta}|^2
  \<\phi^r_{\alpha \beta}|\phi^r_{\alpha \beta}\>
  +\sum_{\alpha\neq\alpha',\beta=\beta'}
  c_{\alpha\beta}c^*_{\alpha'\beta'}
  \<\phi^r_{\alpha'\beta'} |\phi^r_{\alpha \beta}\> \nonumber\\
  &+& \sum_{\alpha=\alpha',\beta\neq\beta'} c_{\alpha\beta} c^*_{\alpha'\beta'}
  \<\phi^r_{\alpha'\beta'}|\phi^r_{\alpha \beta}\> 
  +\sum_{\alpha\neq\alpha',\beta\neq\beta'} c_{\alpha\beta} c^*_{\alpha'\beta'}
  \< \phi^r_{\alpha'\beta'}|\phi^r_{\alpha \beta}\>.
\label{Eqn:Schmidt}
\end{eqnarray}
Now, using the form of $|\phi_{\alpha\beta}^r\>$, and exploiting the
canonical relation one can show that 

\beq
  \<\phi_{\alpha'\beta'}^r|\phi_{\alpha\beta}^r\>=
  \<\alpha'\alpha|R_0\> \<L_0|\beta'\beta\>
  \delta_{\alpha'\alpha}\delta_{\beta' \beta}
  +\sum_s \gamma_s^r \<\alpha'\alpha|R_s\>\<L_s|\beta'\beta\>.
\eeq
Similarly, for $|\phi_{\alpha \beta}^{N-r}\>$, we can write 

\begin{eqnarray}
\< \phi_{\beta' \alpha'}^{N-r}|\phi_{\beta \alpha}^{N-r}\>=\< \beta' \beta|R_0\>\< L_0|\alpha' \alpha\> \delta_{\beta'\beta}\delta_{\alpha' \alpha}+\sum_{s}\gamma_s^{N-r} \< \beta' \beta|R_s\>\< L_s|\alpha' \alpha\>.
\end{eqnarray}
Plugging these in Eq. \eqref{Eqn:Schmidt}, we get

\begin{eqnarray}
\lambda_k^r&=&\sum_{\alpha, \beta} |c_{\alpha\beta}|^2\<\alpha \alpha|R_0\>\< L_0|\beta\beta\>  +\sum_s \gamma_s^{r}  \sum_{\alpha, \beta} |c_{\alpha\beta}|^2\< \alpha \alpha|R_s\>\< L_s|\beta\beta\>\nonumber\\
&+&\sum_{ss'}\gamma_s^r \sum_{\alpha\neq  \alpha', \beta}c_{\alpha\beta} {c^*_{\alpha'\beta}}\< \alpha' \alpha|R_s\>\< L_s|\beta\beta\> \nonumber\\&+&
\sum_{ss'}\gamma_s^r \sum_{\alpha', \beta \neq \beta'}c_{\alpha\beta} {c^*_{\alpha\beta'}}\< \alpha \alpha|R_s\>\< L_s|\beta'\beta\>+
\sum_{ss'}\gamma_s^r \sum_{\alpha \neq   \alpha', \beta \neq \beta'}c_{\alpha \beta} {c^*_{\alpha'\beta'}}\< \alpha' \alpha|R_s\>\< L_s|\beta'\beta\>.
\end{eqnarray}

Similarly, 

\begin{eqnarray}
\Lambda(r,N)_k&=&\sum_{\alpha, \beta}
|d_{\alpha\beta}|^2 \<\alpha\alpha|R_0\> \<L_0|\beta\beta\>
+\sum_s \gamma_s^{N-r}  \sum_{\alpha, \beta} |d_{\alpha\beta}|^2\< \alpha \alpha|R_s\>\< L_s|\beta\beta\>\nonumber\\
&+&\sum_{ss'}\gamma_s^{N-r} \sum_{\alpha\neq  \alpha', \beta}d_{\alpha\beta} {d^*_{\alpha'\beta}}\< \alpha' \alpha|R_s\>\< L_s|\beta\beta\> \nonumber\\&+&
\sum_{ss'}\gamma_s^{N-r} \sum_{\alpha, \beta \neq \beta'}d_{\alpha\beta} {d^*_{\alpha\beta'}}\< \alpha \alpha|R_s\>\< L_s|\beta'\beta\>+
\sum_{ss'}\gamma_s^{N-r} \sum_{\alpha \neq   \alpha', \beta \neq \beta'}d_{\alpha \beta} {d^*_{\alpha'\beta'}}\< \alpha' \alpha|R_s\>\< L_s|\beta'\beta\>.\nonumber\\
\end{eqnarray}

Now if we consider $N \gg1$, we can drop the higher order term and
approximate $\lambda_k^{N-r}$ as 

\begin{eqnarray}
  \lambda_k^{N-r}&=&\sum_{\alpha, \beta}
  |d_{\alpha\beta}|^2\<\alpha\alpha|R_0\>\<L_0|\beta\beta\>.
\end{eqnarray}
Hence, we can write

\begin{eqnarray}
  \lambda_k^N &=& \lambda_k^r \lambda_k^{N-r}=
\(\sum_{\substack{\alpha, \beta}} |c_{\alpha\beta}|^2
\<\alpha\alpha|R_0\> \<L_0|\beta\beta\>\)
\(\sum_{\substack{\alpha,  \beta}}
|d_{\alpha\beta}|^2 \<\alpha\alpha|R_0\> \<L_0|\beta\beta\>\) \nonumber\\
&+&\sum_s \gamma_s^r \( \sum_{\substack{\alpha, \beta}}
|c_{\alpha\beta}|^2 \<\alpha\alpha|R_s\>
\<L_s|\beta\beta\>) 
+\sum_{\substack{\alpha\neq\alpha', \beta}}
c_{\alpha\beta} c^*_{\alpha'\beta} \<\alpha'\alpha|R_s\>\<L_s|\beta\beta\>
\right.\nonumber\\
&+& \left.\sum_{\substack{\alpha, \beta\neq\beta'}}
c_{\alpha\beta} c^*_{\alpha\beta'} \<\alpha\alpha|R_s\>\<L_s|\beta'\beta\>
+ \sum_{\substack{\alpha\neq\alpha', \beta\neq\beta'}}
c_{\alpha\beta} c^*_{\alpha'\beta'} \<\alpha'\alpha|R_s\>
\<L_s|\beta'\beta\>\) \( \sum_{\substack{\alpha=\alpha',\beta=\beta'}}
|d_{\alpha\beta}|^2 \<\alpha\alpha|R_0\> \<L_0|\beta\beta\>)\),\nonumber\\
&=&\Lambda_k+\sum_s \gamma_s^r \Lambda^s_k,
\end{eqnarray}
with 

\begin{eqnarray}
\Lambda_k&=&\(\sum_{\substack{\alpha,   \beta}}
|c_{\alpha\beta}|^2\< \alpha \alpha|R_0\>\< L_0|\beta\beta\>\)
\(\sum_{\substack{\alpha, \beta}} |d_{\alpha\beta}|^2\< \alpha
\alpha|R_0\>\< L_0|\beta\beta\>\),\\ 
\Lambda_k^s &=&\(\sum_{\substack{\alpha, \beta}}
|c_{\alpha\beta}|^2\< \alpha \alpha|R_s\>
\<L_s|\beta\beta\>+\sum_{\substack{\alpha\neq\alpha', \beta}}
c_{\alpha\beta} c^*_{\alpha'\beta} \<\alpha'\alpha|R_s\>
\<L_s|\beta\beta\>
+\sum_{\substack{\alpha, \beta\neq\beta'}}
c_{\alpha\beta} c^*_{\alpha\beta'} \<\alpha\alpha|R_s\>
\<L_s|\beta'\beta\> \right.\nonumber\\
&+&\left.\sum_{\substack{\alpha\neq\alpha',\\\beta\neq\beta'}}
c_{\alpha \beta} c^*_{\alpha'\beta'}
\<\alpha'\alpha|R_s\>\<L_s|\beta'\beta\>\)
\(\sum_{\substack{\alpha, \beta}}
|d_{\alpha\beta}|^2 \<\alpha\alpha|R_0\> \<L_0|\beta\beta\>\).
\end{eqnarray}
Now if we plug  $\lambda_k^N$ in the entropy equation as given in  Eq. (\ref{eqn:entropy}) in the main text, we get
\begin{eqnarray}
  S(r) &=& -\sum_k  \Lambda_k \log \Lambda_k
  - \sum_s \gamma_s^r \[\Lambda_k^s +
  \sum_k  \Lambda_k^s\log\Lambda_k \]
  - \sum_{ss'} \gamma_s^r \gamma_{s'}^r
  \frac{\Lambda_k^s \Lambda_k^{s'}}{\Lambda_k},\nonumber\\
  &=& S_0 - \sum_s \gamma_s^r \[\sum_k \Lambda_k^s 
  + \sum_k  \Lambda_k^s \log \Lambda_k \]
  - \sum_{ss'} \gamma_s^r \gamma_{s'}^r \sum_k
  \frac{\Lambda_k^s \Lambda_k^{s'}}{\Lambda_k}. \nonumber\\
\end{eqnarray}


\end{document}